\documentclass[conference]{IEEEtran}
\IEEEoverridecommandlockouts
\usepackage{cite}
\usepackage{amsmath,amssymb,amsfonts}
\usepackage{graphicx}
\usepackage{textcomp}
\usepackage{xcolor}

\usepackage{algorithm}
\usepackage[noend]{algpseudocode}
\usepackage{mathrsfs}
\usepackage[tight,footnotesize]{subfigure}
\usepackage{url}
\usepackage{amsthm}
\newcommand{\verde}{{\it {ImVerde}}}
\newtheorem{theorem}{Theorem}
\newtheorem{lemma}[theorem]{Lemma}
\newtheorem{corollary}[theorem]{Corollary}
\newtheorem{problem}{Problem}
\usepackage{array,graphicx,subfig}
\usepackage{multirow,color}

\def\BibTeX{{\rm B\kern-.05em{\sc i\kern-.025em b}\kern-.08em
    T\kern-.1667em\lower.7ex\hbox{E}\kern-.125emX}}
\begin{document}

\title{{\it ImVerde}: Vertex-Diminished Random Walk for Learning Imbalanced Network Representation}

\author{\IEEEauthorblockN{Jun Wu}
\IEEEauthorblockA{\textit{Arizona State University} \\
junwu6@asu.edu}
\and
\IEEEauthorblockN{Jingrui He}
\IEEEauthorblockA{\textit{Arizona State University} \\
jingrui.he@asu.edu}
\and
\IEEEauthorblockN{Yongming Liu}
\IEEEauthorblockA{\textit{Arizona State University} \\
Yongming.Liu@asu.edu}
}

\maketitle

\begin{abstract}
Imbalanced data widely exist in many high-impact applications. An example is in air traffic control, where among all three types of accident causes, historical accident reports with `personnel issues' are much more than the other two types (`aircraft issues' and `environmental issues') combined. Thus, the resulting data set of accident reports is highly imbalanced. On the other hand, this data set can be naturally modeled as a network, with each node representing an accident report, and each edge indicating the similarity of a pair of accident reports. Up until now, most existing work on imbalanced data analysis focused on the classification setting, and very little is devoted to learning the node representations for imbalanced networks. To bridge this gap, in this paper, we first propose Vertex-Diminished Random Walk (VDRW) for imbalanced network analysis. It is significantly different from the existing Vertex Reinforced Random Walk by discouraging the random particle to return to the nodes that have already been visited. This design is particularly suitable for imbalanced networks as the random particle is more likely to visit the nodes from the same class, which is a desired property for learning node representations.  Furthermore, based on VDRW, we propose a semi-supervised network representation learning framework named \verde\ for imbalanced networks, where context sampling uses VDRW and the limited label information to create node-context pairs, and balanced-batch sampling adopts a simple under-sampling method to balance these pairs from different classes. Experimental results demonstrate that \verde\ based on VDRW outperforms state-of-the-art algorithms for learning network representations from imbalanced data.
\end{abstract}


\section{Introduction}
Nowadays, big data is being generated across multiple high impact application domains. Often times, the target data set is imbalanced in terms of the proportion of examples from various classes. For example, in air traffic control, the large number of historical flight accident reports can be used to analyze and identify the leading indicators of various accident causes. According to National Transportation Safety Board\footnote{\url{https://www.ntsb.gov/Pages/default.aspx}}, there are three major types of accident causes, including aircraft issues, personnel issues, and environmental issues. Historical records with `personnel issues' are much more than the other two types combined, thus creating an imbalanced data set. For example, of the accident reports between aircraft control and directional control, aircraft control of them are due to `personnel issues', which dominate the entire data set. On the other hand, such a data set can be naturally modeled as a network, where each node corresponds to a historical accident report, and each edge reflects the similarity between two reports.

\begin{figure}
\centering
\subfigure[Planetoid]{
\begin{minipage}{0.43\linewidth}
\centering
\includegraphics[width=1.8in]{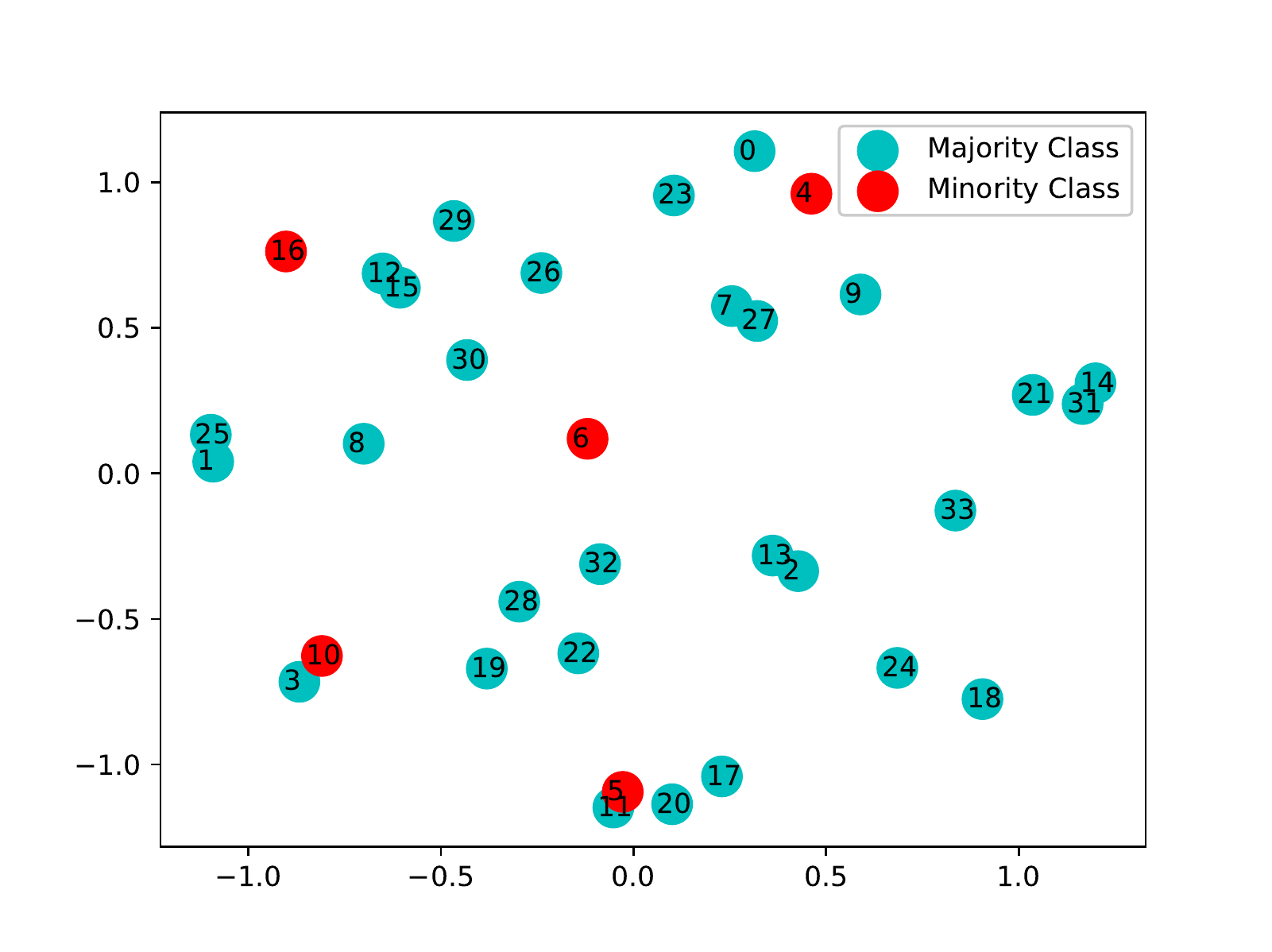}
\end{minipage}
}
\subfigure[\verde{}]{
\begin{minipage}{0.5\linewidth}
\centering
\includegraphics[width=1.8in]{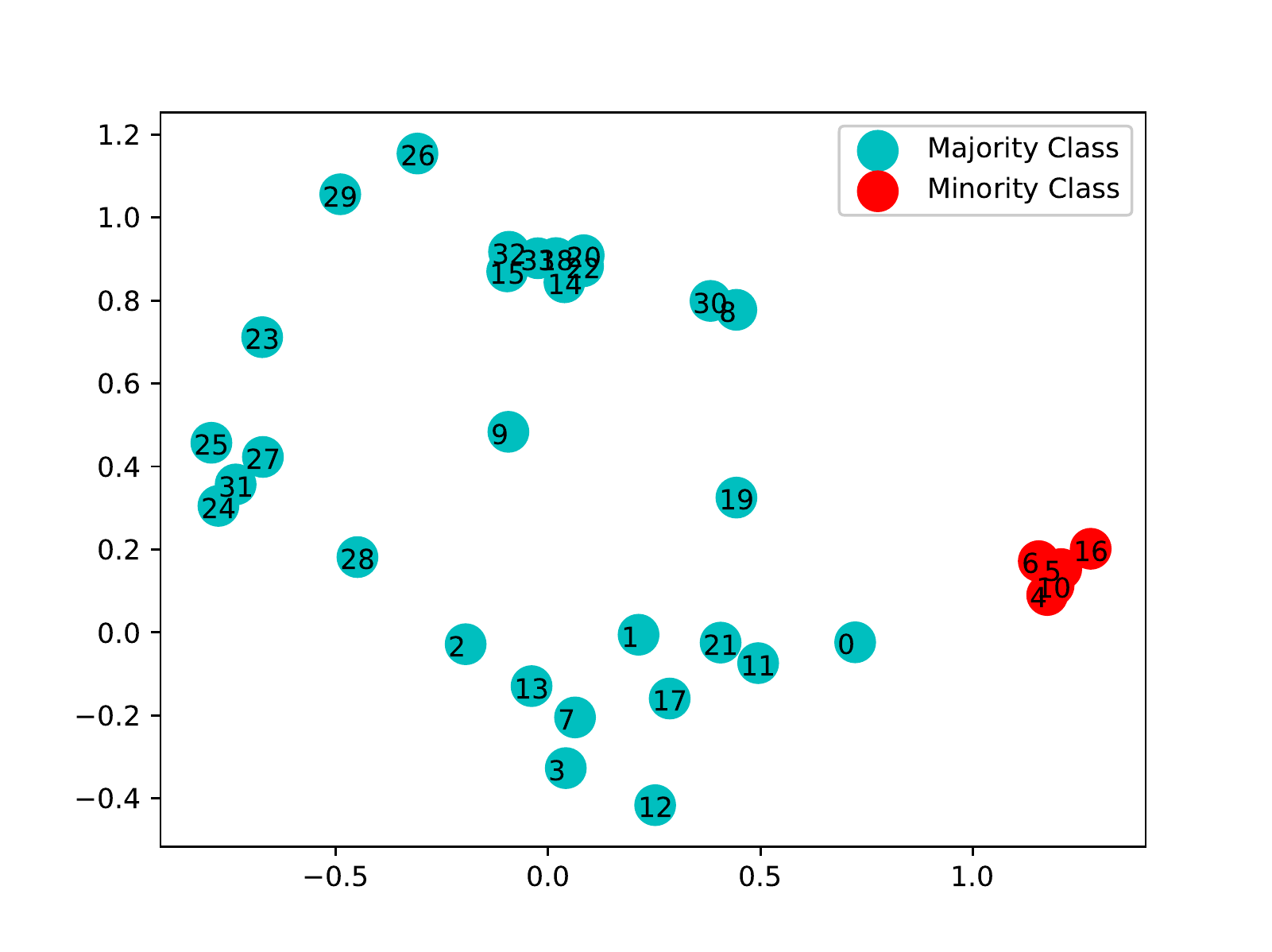}
\end{minipage}
}
\caption{Network embedding in 2-dimensional space with two imbalanced classes from Zachary's Karate network~\cite{zachary1977information} (shown in Figure \ref{fig:vis}(a)) using: (a) Planetoid~\cite{yang2016revisiting} (b) proposed \verde{}. (t-SNE~\cite{maaten2008visualizing} is adopted to visualize the network embedding; best viewed in color)}\label{fig:imbalance_ex}
\vspace{-6mm}
\end{figure}

Despite the extensive existing work on imbalanced classification~\cite{he2009learning,chawla2002smote,huang2016learning}, up until now, very little (if any) effort has been devoted to learning the node representation from imbalanced networks. Most existing work on network embedding either works in the unsupervised fashion, i.e., not considering the label information at all, or is implicitly assuming a balanced setting, i.e., not suitable for the imbalanced setting. More specifically, unsupervised embedding approaches~\cite{hamilton2017representation, perozzi2014deepwalk, lai2017prune, duran2017learning} aimed to learn the node representations which preserve the graph topological structure in the feature space; and
semi-supervised methods~\cite{yang2016revisiting,liang2018semi,tang2015pte} implicitly assumed that the labeled set consisted of roughly equal number of examples from each class. However, in the presence of imbalanced labeled set, these methods will inevitably suffer from imbalanced node-context pairs, resulting in node representations not amiable to the following classification task.

To bridge this gap, in this paper we propose a generic framework named \verde{} for learning node representations from imbalanced networks. It is based on a novel random walk model named Vertex-Diminished Random Walk (VDRW), where the basic idea is that the transition probability to one node decreases each time it is visited by the random particle. Notice that when learning the embedding vectors from imbalanced networks, compared with the existing Vertex Reinforced Random Walk~\cite{pemantle1992vertex}, VDRW can capture the node-context pairs from the minority class with a higher accuracy.

Furthermore, we sample the context using both graph structure and label information. In order to extract node-context pairs using the two types of information jointly, we propose a jumping scheme to combine them into one unified framework. In other words, if a node is labeled, it would have a certain probability jumping to other nodes with the same labels in one step rather than walking randomly to neighbors. Finally, in order to balance the extracted node-context pairs in imbalanced networks, we use a simple under-sampling method for mini-batch sampling. Figure \ref{fig:imbalance_ex} shows that compared with Planetoid~\cite{yang2016revisiting}, the network embedding produced by the proposed \verde\ leads to better separability between the two classes in the imbalanced data set. To validate the effectiveness of the proposed network representation framework, we conduct experiments on several real network data sets with promising results.

The main contributions of this paper can be summarized as follows:
\begin{itemize}
\item We propose a novel random walk model, named Vertex-Diminished Random Walk (VDRW), which adjusts the transition probability each time a node is visited by the random particle. It leads to an effective context sampling method for imbalanced networks, and its convergence properties are analyzed.
\item Based on VDRW, we propose a semi-supervised network representation framework named \verde, in which two strategies are introduced to improve the quality of node-context pairs: \emph{\textbf{Context Sampling}} and \emph{\textbf{Balanced-batch Sampling}}. \emph{\textbf{Context Sampling}} extracts node-context pairs based on both graph structure and label information, and \emph{\textbf{Balanced-batch Sampling}} aims to keep the extracted pairs balanced.
\item The proposed framework is evaluated on several data sets, with experimental results demonstrating its superior performance over state-of-the-art techniques. These results are also consistent with our analysis on imbalanced data in terms of, e.g., high-quality node-context pairs, separability of different classes, etc. 
\end{itemize}

The rest of the paper is organized as follows. We review the related work in Section II. VDRW is introduced in Section III and Section IV presents our proposed semi-supervised network representation framework. The extensive experimental results and discussion are provided in Section V. Finally we conclude the paper in Section VI.

\section{Related Work}
In this section, we briefly review the related work on imbalanced data analysis and network representation learning.

\subsection{Imbalanced Data Analysis}
Imbalanced data mining has attracted significant attention in many high-impact applications since standard algorithms implicitly assume the balanced distribution in data and may
fail to analyze class-imbalanced data~\cite{lopez2013insight}. And in most cases, the data annotated by workers~\cite{zhou2016crowdsourcing,zhou2017multic2} hardly ensure the balanced distribution.
Up until now, over-sampling~\cite{chawla2002smote,he2008adasyn} and under-sampling~\cite{tomek1976two} are two mainstream approaches to balance the instance distribution in different classes via adding replicated instances from minority class or reducing instances from majority class. To solve the over-fitting in over-sampling, SMOTE~\cite{chawla2002smote} created synthetic samples between each minority instance and its nearest neighbors. Tomek links~\cite{tomek1976two} were defined as a pair of samples in opposite classes with minimal distance. By removing or cleaning the overlapping instances, Tomek links usually integrated with other sampling methods to target imbalanced problems. Instead of altering data distribution, cost-sensitive methods~\cite{elkan2001foundations} considered the penalty for misclassifying instances to another class. Specifically, misclassified minority instances bear heavier penalty in order to alleviate imbalanced data distribution. Recently deep learning models~\cite{huang2016learning,khan2017cost} are also adopted to deal with imbalanced data. Additionally, minority classes or rare categories normally play important roles in the database, such as financial fraud in bank systems. 
Rare category detection algorithms~\cite{he2008graph,zhou2015muvir,zhou2018sparc,zhou2015rare} aimed to identify those special minority groups from imbalanced data. Compared with the existing work on imbalanced data analysis, in this paper, we study a novel problem setting of learning network representation from imbalanced networks, whereas existing work mainly focuses on the classification problem or the active learning setting.

\subsection{Network Representation Learning}
The objective of network representation is to learn a low-dimensional dense vector for each node in the network. One of the common assumptions in most of the existing work~\cite{wang2017community,hamilton2017representation,cui2018survey,huang2017label} is that nearby nodes have similar embeddings in the feature space. Among these algorithms, there are mainly three categories: factorization-based, random walk based, and CNN-based. The factorization-based algorithms can be traced back to classic dimensionality reduction approaches LLE~\cite{roweis2000nonlinear} and ISOMAP~\cite{tenenbaum2000global}, which preserved the manifold properties in the low-dimensional space. Recently, to capture high-order proximities relationship among the nodes, LINE~\cite{tang2015line} and GraRep~\cite{cao2015grarep} were proposed to learn network representation unsupervisedly. For random walk based models, DeepWalk~\cite{perozzi2014deepwalk} firstly showed that truncated random walk follows the similar power-law distribution as skip-gram model does in word frequency. Subsequently, many (unsupervised or semi-supervised) random work based approaches~\cite{yang2016revisiting,liang2018semi,grover2016node2vec,yang2015network} were proposed to learn node representation from homogeneous networks. And Qiu et al.~\cite{qiu2018network} proposed that some of those embedding approaches, e.g., DeepWalk~\cite{perozzi2014deepwalk} and node2vec~\cite{grover2016node2vec}, can be transformed into the matrix factorization problems.
In addition, motivated by convolutional neural network architecture,
Graph Convolutional Networks (GCN)~\cite{kipf2016semi,defferrard2016convolutional,chen2018fastgcn} extended convolution operation to graph-structured data using spectral graph theory for semi-supervised node classification. Similarly, GraphSAGE~\cite{hamilton2017inductive} aimed to learn a function that aggregates embeddings from nodes' local neighborhood for inductive representation learning. To capture the highly non-linear network structure, SDNE~\cite{wang2016structural} used a semi-supervised deep neural network model with non-linear activation functions to preserve first- and second-order proximities among the nodes. However, very little effort has been devoted to learning the node representation from imbalanced networks, which will result in imbalanced node-context pairs. Our proposed \verde\ framework is designed to address this issue through VDRW, a novel context sampling method, as well as the balanced-batch sampling method.

\section{Vertex-Diminished Random Walk}
\label{Vertex-Diminished Random Walk}
In this section, we present the Vertex-Diminished Random Walk (VDRW) for imbalanced network analysis. It resembles the existing Vertex Reinforced Random Walk (VRRW)~\cite{pemantle1992vertex} in terms of the dynamic nature of the transition probabilities, as well as some convergence properties. However, it is fundamentally different from VRRW in which the transition probabilities to one node increase each time it is visited. As our analysis will show, VDRW is particularly suitable for analyzing imbalanced networks as it encourages the random particle to walk within the same class, which in turn will lead to separable node representations in the embedding space between the majority and the minority classes in the imbalanced data set.

\subsection{Notation}
Suppose that an undirected (or directed) network is denoted $G = (V, E)$ where $V$ is the node set consisting of $n$ nodes, and $E$ is the edge set consisting of $m$ edges. Each edge $e =(v_i, v_j) \in E$ is associated with a positive weight $w_{ij} > 0$ if $v_i$ and $v_j$ are connected in the network. Otherwise, $w_{ij} = 0$. For this network, let $\mathbf{R}$ denote the non-negative $n\times n$ transition matrix, whose element in the $i^{\textrm{th}}$ row and $j^{\textrm{th}}$ column is $\mathbf{R}_{i,j}=\frac{w_{ij}}{\sum_jw_{ij}}$. Let $Y_{t}\in\{1,\ldots,n\}$ be the state of the random particle at step $t$ and $\mathbf{S}(t) = [\mathbf{S}_1(t), \cdots, \mathbf{S}_n(t)]\in \mathbb{R}^n$ denote the number of visits to each node up to time $t$ where $\mathbf{S}_i(0) = 0$ for $i=1,\cdots,n$. That is, $\mathbf{S}_i(t+1) = \mathbf{S}_i(t) + \delta_{Y_{t+1},i}$ where $\delta_{Y_{t+1},i} = 1$ if $v_i$ is visited at step $t+1$, namely $Y_{t+1} = i$; otherwise, $\delta_{Y_{t+1},i} = 0$.

\subsection{VDRW: A Novel Random Walk Model}
As illustrated in DeepWalk~\cite{perozzi2014deepwalk}, a short random walk follows a power-law distribution as word frequency in language modeling. By implicitly assuming the balanced setting in labeled training examples~\cite{liang2018semi,yang2016revisiting}, most nodes can find a path within the same class using short random walk. Nevertheless, random walk does not necessarily work for forcing nodes to walk within the same class on an imbalanced network because random walk might follow the node-degree distribution to create walking path. That is, the nodes in minority class are likely to walk to majority class instead of walking within the same class since node degrees of majority class are higher than that of minority class in most cases. As a result, there is not much difference between the nodes coming from the majority and minority classes in terms of their walking paths, leading to non-separable representations in the embedding space.
\begin{figure}[htp]
\centering
\includegraphics[width=3.3in]{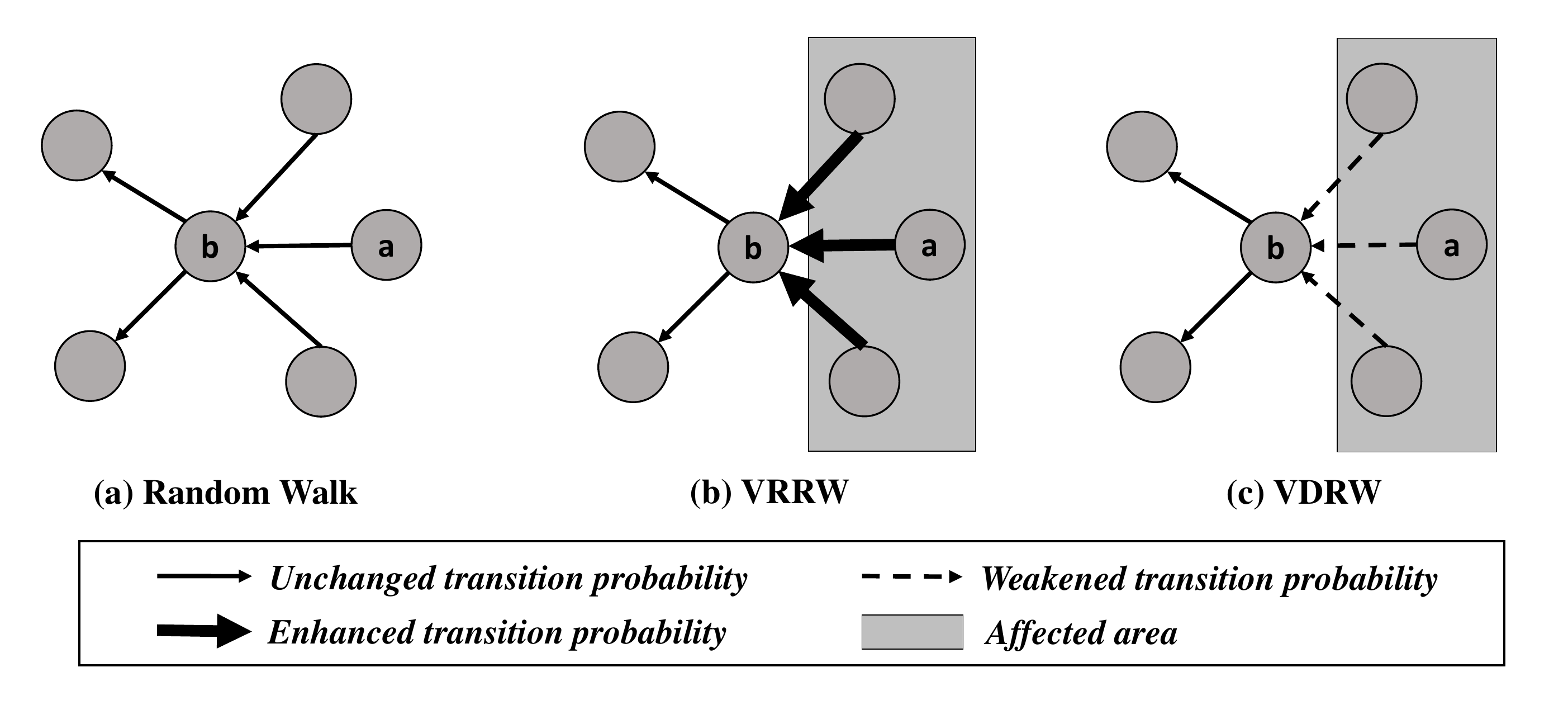}
\caption{Comparison among Random Walk, VRRW~\cite{pemantle1992vertex} and VDRW}\label{random_walk}
\vspace{-6mm}
\end{figure}

\begin{figure*}
\centering
\subfigure[Karate Network]{
\begin{minipage}{4cm}
\centering
\includegraphics[width=4.5cm]{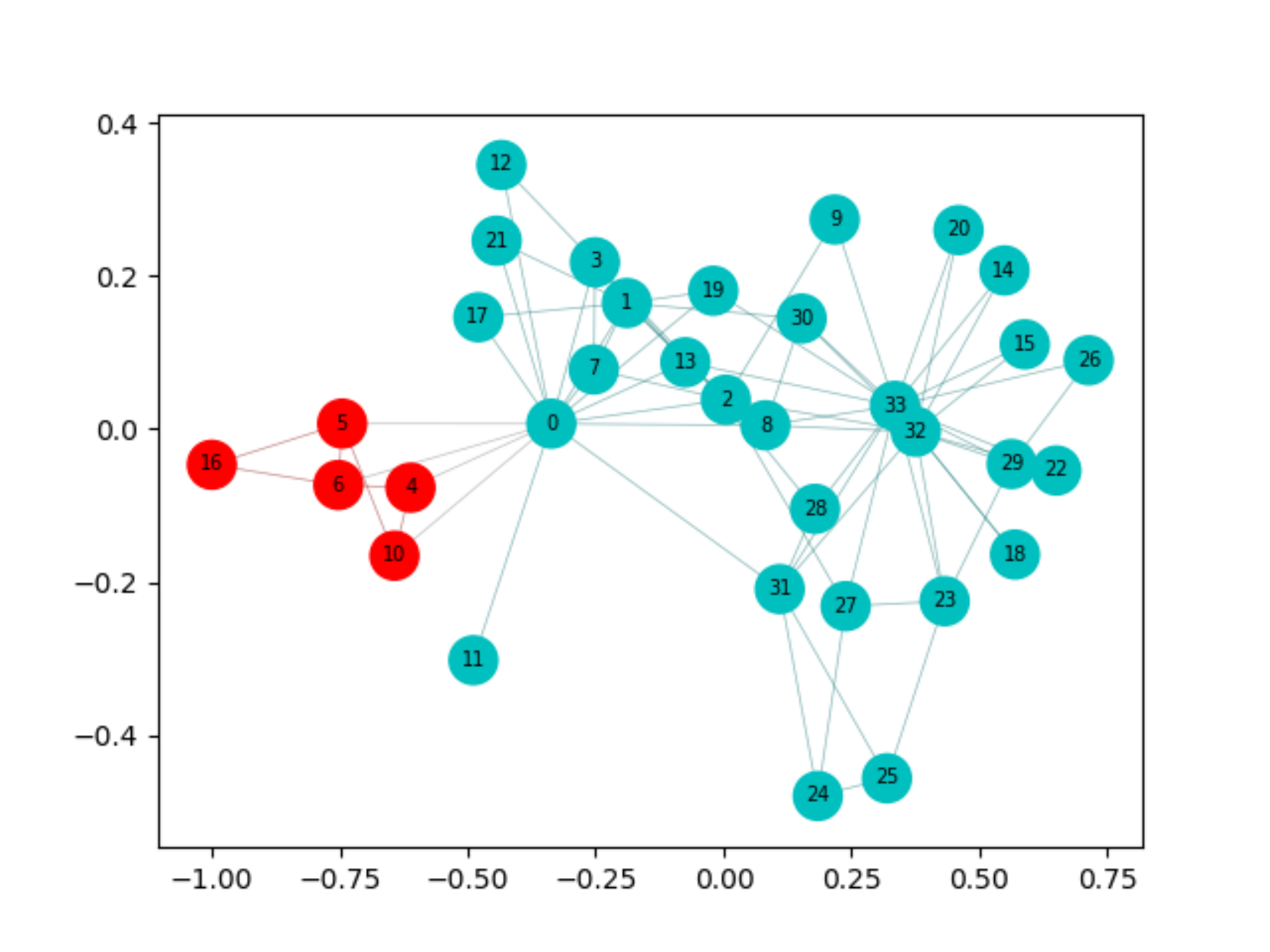}
\end{minipage}
}
\subfigure[Random Walk]{
\begin{minipage}{4cm}
\centering
\includegraphics[width=4.5cm]{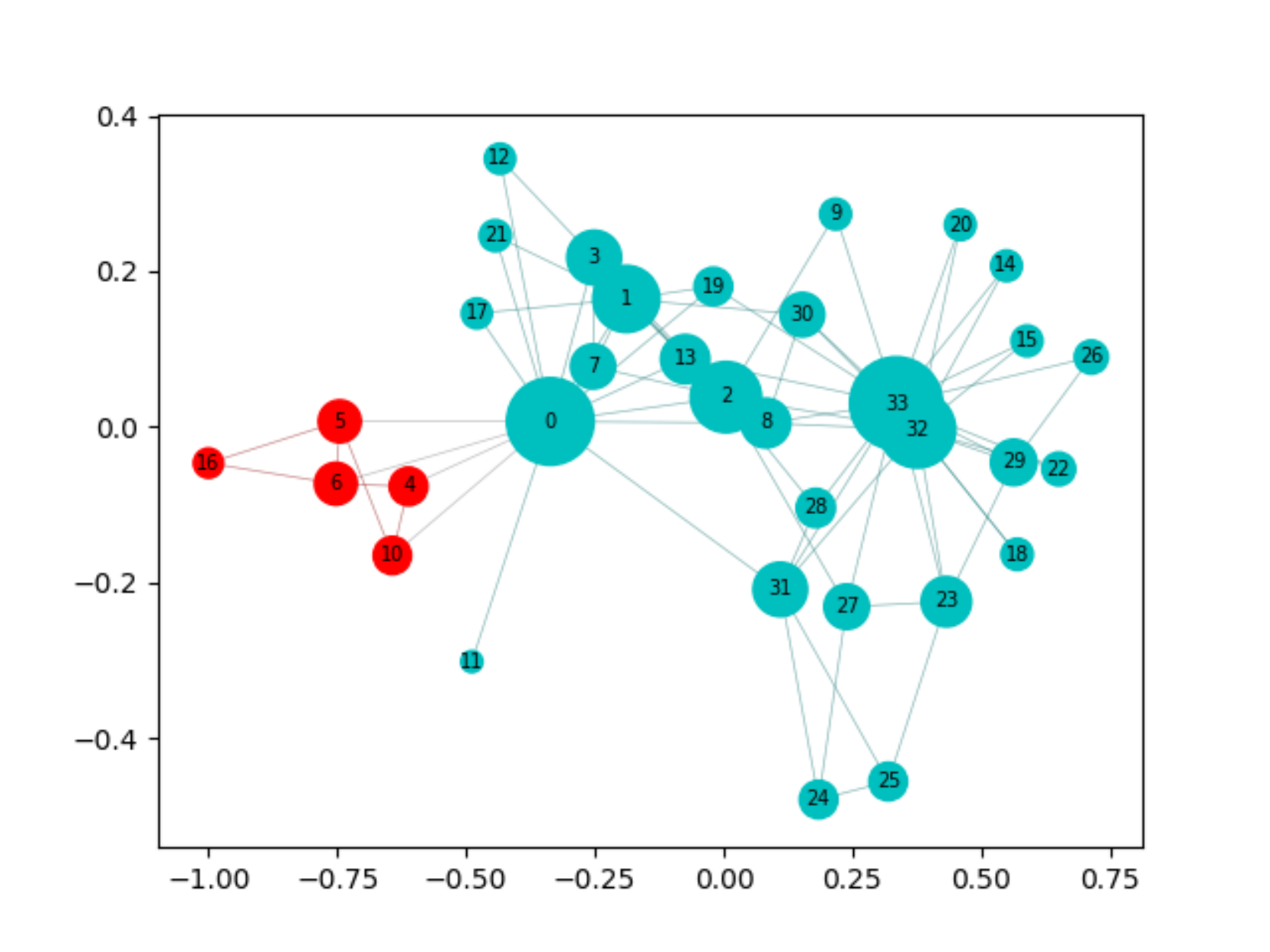}
\end{minipage}
}
\subfigure[VRRW]{
\begin{minipage}{4cm}
\centering
\includegraphics[width=4.5cm]{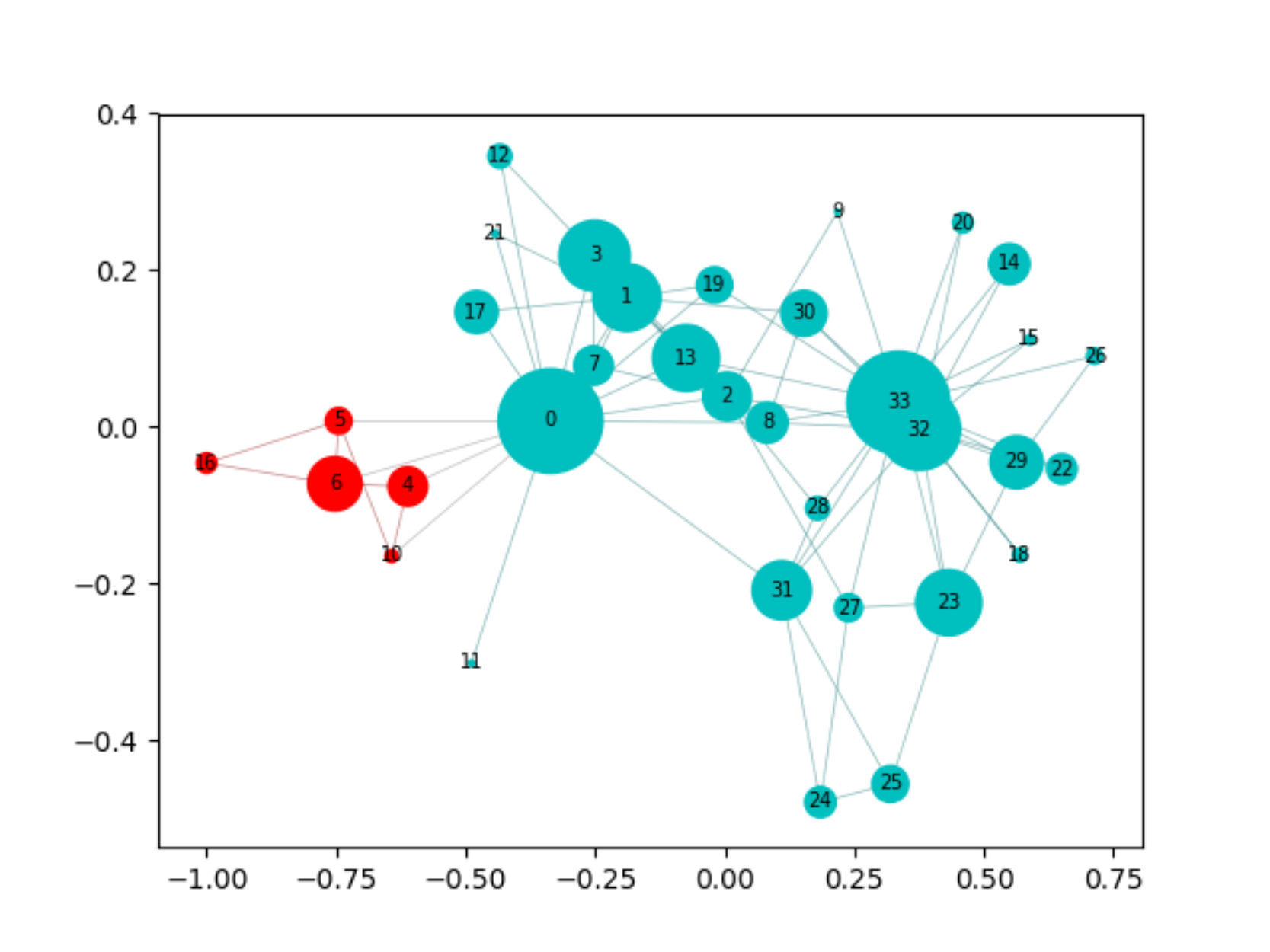}
\end{minipage}
}
\subfigure[VDRW]{
\begin{minipage}{4cm}
\centering
\includegraphics[width=4.5cm]{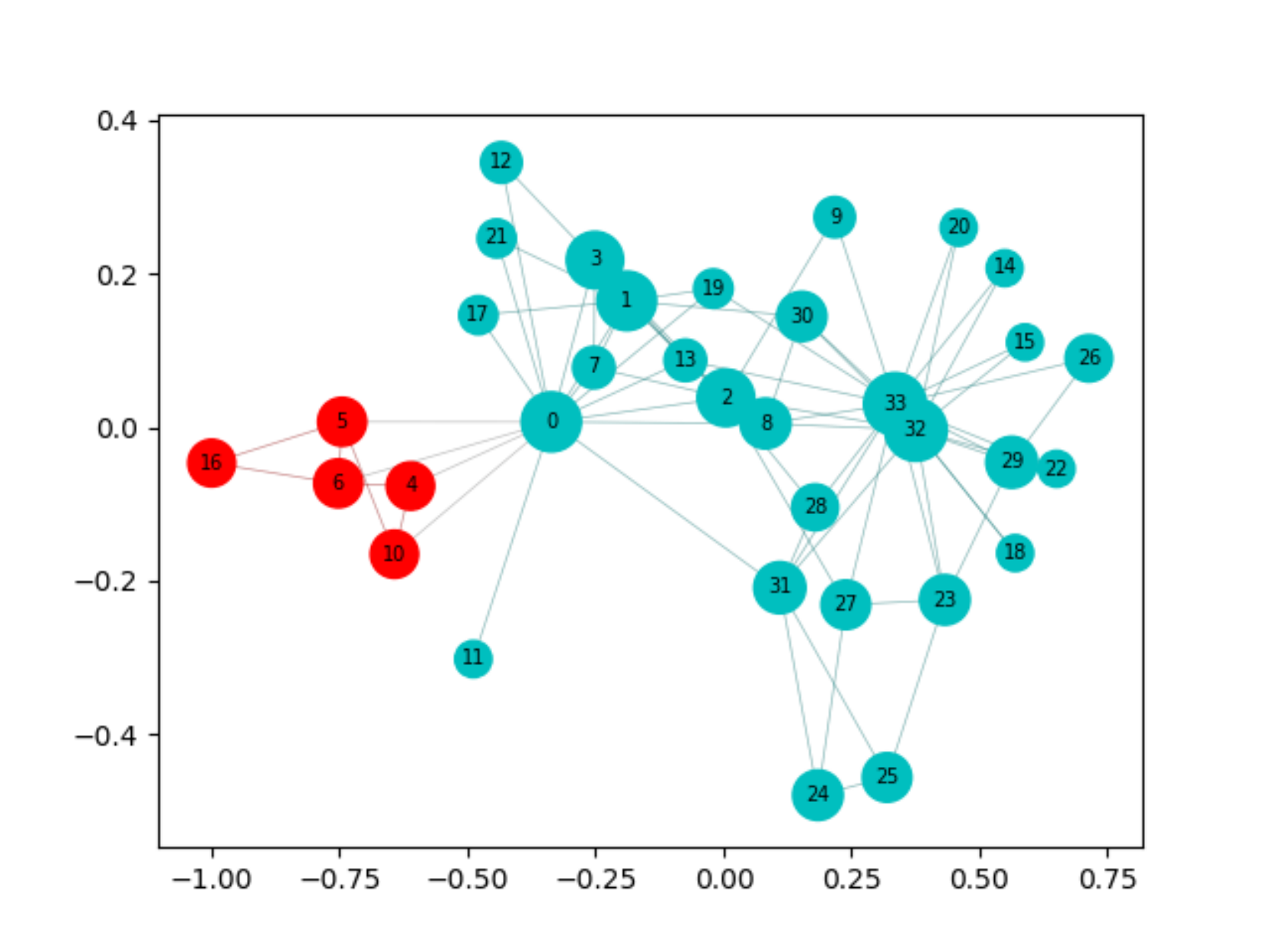}
\end{minipage}
}
\caption{The node visited frequency distribution after walking 2000 steps on Zachary's Karate network~\cite{zachary1977information} (The node size corresponds to the visited times and the colors corresponds to different node classes)}\label{fig:vis}
\vspace{-6mm}
\end{figure*}

To address this problem, in this paper, we propose a novel random walk model named Vertex-Diminished Random Walk (VDRW) based on the idea that the transition probability to one node decreases each time it is visited. It resembles the existing Vertex-Reinforced Random Walk (VRRW)~\cite{pemantle1992vertex} in terms of the dynamic nature of the transition probabilities. But the transition probabilities to visit one node increase each time it is visited in VRRW. Figure~\ref{random_walk} illustrates the difference among random walk, VRRW, and VDRW in a directed graph. When node $b$ is visited by node $a$, there is no change regarding the transition probability in random walk. However, all the transitions to node $b$ are affected in VRRW and VDRW when it is visited by node $a$. Those transitions would be weakened in VDRW, whereas VRRW forces them to be enhanced after one node is visited each time. Later, we will explain why VDRW is adopted in imbalanced network analysis instead of random walk or VRRW.

Now we first present a generalized random walk in which the transition probability changes when one node is visited by the random particle.
We introduce a visiting function $f(\mathbf{S}_i(t))$ to capture the change behavior of the transition probabilities with respect to visited times $\mathbf{S}_i(t)$ for node $v_i$. Therefore, after moving for $t$ steps, $f(\mathbf{S}_i(t))$ keeps track of how the transition probabilities to $v_i$ change. In general, the transition probability to $v_j$ at time $t+1$ is defined as follows:
\begin{equation}
P(Y_{t+1} = j|\mathscr{F}_t) = \frac{\mathbf{R}_{Y_t,j}f(\mathbf{S}_j(t))}{\sum_i \mathbf{R}_{Y_t,i}f(\mathbf{S}_i(t))}
\end{equation}
where $\mathscr{F}_t$ is the $\sigma$-field\footnote{$\sigma$-field on a set $X$ is a collection of the whole subsets of $X$ that includes the empty subset.} generated by $\{Y_k\}_{1 \leq k \leq t}$. The visiting function $f(\mathbf{S}_i(t))$ determines how the transition probability changes over time. Obviously, it has two special cases: $f(\mathbf{S}_i(t)) = 1$ for random walk, and $f(\mathbf{S}_i(t)) = \mathbf{S}_i(t) + 1$ for VRRW. In other words, the transition probability is always constant in traditional random walk, whereas linearly relying on the visited times of one node in VRRW.

VDRW is a stochastic process in which the transition probability to one node is decreased each time it is visited, resulting in more likely to explore the unvisited nodes compared to random walk or VRRW. It is inversely proportional to the visited times $\mathbf{S}_i(t)$ at time $t$. In this paper, we define the visiting function $f(\mathbf{S}_i(t))$ for VDRW as follows:
\begin{equation}
\label{2}
f(\mathbf{S}_i(t)) = \alpha^{\mathbf{S}_i(t)}
\end{equation}
where $0 < \alpha < 1$ is a specific hyperparameter. Intuitively, when node $v_i$ is visited, the transition to $v_i$ from all other nodes will be diminished as shown in Figure \ref{random_walk}(c). 

Of importance of this algorithm is to change the transition probabilities with respect to the number of times one node has been visited. Nevertheless, it is fundamentally different from VRRW. In VRRW, the transition probabilities to one node increase each time it is visited. High-degree nodes would be visited frequently because the transition probabilities to those nodes are more likely to be increased. On the contrary, VDRW encourages the random particle to explore the unvisited or less-frequently visited nodes in the network. That is, the visited frequency distribution using VDRW would be denser than that using VRRW or random walk. Figure \ref{fig:vis} is the visited frequency distribution for all the nodes on Zachary's Karate network~\cite{zachary1977information} after walking 2000 steps (best viewed in color). It can be observed that the visited frequency in random walk follows the node degree distribution on the original graph. And it is much sparser in VRRW than random walk because VRRW implicitly forces the high-degree nodes to be easily visited by gradually increasing the transition probabilities to those nodes. On the contrary, VDRW would encourage all the nodes to be visited smoothly as shown in Figure \ref{fig:vis}(d). Besides, the experiments in Section \ref{case_study} demonstrate that the nodes tend to walk within the same class using short VDRW.

The intuition of adopting VDRW in our network embedding framework can be simply explained as follow. Short VDRW decreases the transition probabilities to one node when it is visited each time, leading to explore unvisited or less-frequently visited nodes instead of high-degreed nodes. That is, nodes in the imbalanced network would have smooth visited frequency distribution. And the transition to the majority class would be more likely to be decreased. That forced the nodes in the minority class to walk within the same class rather than explore the nodes in the majority class. Thus high-quality node-context pairs can be extracted for our network representation framework using short VDRW.

\subsection{Convergence Analysis}
In this subsection, we analyze the convergence properties of the function $\mathbf{V}(t)\in \mathbb{R}^n$ with respect to step $t$, in which the $i$-th element is defined as $\mathbf{V}_i(t) = f(\mathbf{S}_i(t))/\sum_if(\mathbf{S}_i(t))$. Obviously, $\mathbf{V}(t)$ belongs to the $(n-1)$-simplex\footnote{$k$-simplex is defined as a $k$-dimensional polytope with $k + 1$ vertices.} $\Delta \subseteq \mathbb{R}^n$. To analyze the convergence of VDRW, we use the same definitions as~\cite{pemantle1992vertex}. For any vector $\mathbf{v} \in \Delta$ with $H(\mathbf{v}) = \sum_{ij}\mathbf{R}_{ij}{v}_i{v}_j > 0$, Markov transition matrix $\mathbf{M} \in \mathbb{R}^{n\times n}$ is defined as:
\begin{equation}
\mathbf{M}_{ij}(\mathbf{v}) = \frac{\mathbf{R}_{ij}{v}_j}{\sum_{j}\mathbf{R}_{ik}{v}_k}
\end{equation}
And let $\mathcal{C} \subseteq \Delta$ be the set of points $\mathbf{v}$ with $\pi(\mathbf{v}) = \mathbf{v}$ where the vector $\pi(\mathbf{v}) \in \Delta$ is defined as:
\begin{equation}
\pi_i(\mathbf{v}) = \frac{\sum_{j}\mathbf{R}_{ij}{v}_i{v}_j}{H(\mathbf{v})}
\end{equation}
Let $\mathcal{C}_0 \subseteq \Delta$ be the set of points $\mathbf{v}$ for which $\mathbf{M(v)}$ is reducible. The following theorem provides our main results regarding the convergence of VDRW.

\begin{theorem}
With probability one, dist $(\mathbf{V}(t), \mathcal{C} \cup \mathcal{C}_0) \longrightarrow 0$ where dist $(x, A) = \inf\{|x-y|: y \in A\}$.
\end{theorem}

To prove this theorem, we introduce the following lemma
regarding the convergence conditions of VDRW. When the defined visiting function satisfies this lemma, it can be proven that Theorem 1 holds for VDRW.

More specifically, for any $t$, let $\mathbf{M}_t(t), \mathbf{M}_t(t+1), \cdots$ denote a Markov chain with fixed transition matrix $\mathbf{M}(\mathbf{V}(t))$ beginning at $Y_t$ at time $t$, $\mathbf{S'}(t) = \mathbf{S}(t)$ and $\mathbf{S'}(i) = \mathbf{S'}(i-1) + e_{\mathbf{M}_t(i)}$ for $i>t$ where $e_j$ denotes the $j$-th standard basis vector. Let $\mathcal{N}$ be a closed subset of a simplex with $\mathcal{N} \cap (\mathcal{C} \cup \mathcal{C}_0) = \emptyset$.

\begin{lemma}
If $\{f\big(\mathbf{S'}(i)\big)\}$ has Markov property, and $\big(f\big(\mathbf{S'}(t+L)\big) - f\big(\mathbf{S'}(t)\big)\big)\big/\big(\sum_if\big(\mathbf{S'}_i(t+L)\big) - \sum_if\big(\mathbf{S'_i}(t)\big)\big)$ approaches a point-mass at $\pi(\mathbf{v})$ in distribution as $L$ increases, then there exists $c>0$ such that $\mathbf{E}\big(H(\mathbf{V}(n+L))|\mathbf{V}(n)\big) > H\big(\mathbf{V}(n)\big) + c/n$ whenever $\mathbf{V}(n) \in \mathcal{N}$.
\end{lemma}
\begin{proof}
 The detailed proof is omitted here for brevity.
\end{proof}

When $f(\mathbf{S}(t)) = 1 + \log \alpha \cdot \mathbf{S}(t)$, we have the following result.
\begin{equation*}
\begin{aligned}
&\frac{f(\mathbf{S'}(t+L)) - f(\mathbf{S'}(t))}{\sum_if(\mathbf{S'_i}(t+L)) - \sum_if(\mathbf{S'_i}(t))} \\
&= \frac{(1+ \log\alpha \cdot {\mathbf{S'}(t+L)}) - (1+ \log\alpha \cdot {\mathbf{S'}(t)})}{(n - \log \alpha \cdot \sum_i \mathbf{S'_i}(t+L)) -  (n - \log \alpha \cdot \sum_i \mathbf{S'_i}(t))} \\
&= \frac{\log \alpha \cdot (\mathbf{S'}(t+L) - \mathbf{S'}(t))}{\log \alpha \cdot L} \\
&= \frac{\mathbf{S'}(t+L) - \mathbf{S'}(t)}{L}
\end{aligned}
\end{equation*}
As shown in \cite{pemantle1992vertex}, $(\mathbf{S'}(t+L) - \mathbf{S'}(t))/L$ approaches a point-mass at $\pi(\mathbf{v})$ in distribution as $L$ increases. Therefore, $f(\mathbf{S}(t)) = 1 + \log \alpha \cdot \mathbf{S}(t)$ meets the basic condition in the Lemma 2 above. It is obvious that the function $f(\mathbf{S}(t))$ defined in Equation~(\ref{2}) has the following Taylor expansion:
\begin{equation}
f(\mathbf{S}(t)) = \alpha^{\mathbf{S}(t)} = 1 + \log \alpha \cdot \mathbf{S}(t) + O(\log \alpha \cdot \mathbf{S}(t))
\end{equation}
Thus, we use $f(\mathbf{S}(t)) = \alpha^{\mathbf{S}(t)}$ to define the visiting function in VDRW. Moreover, the following Corollary 3 presents the conditions under which the convergence of $\mathbf{V}(t)$ is almost surely.

\begin{corollary}
If all the off-diagonal entries of $\mathbf{R}$ are positive and all the principal minors of $\mathbf{R}$ are invertible, $\mathbf{V}(t)$ converges almost surely.
\end{corollary}
\begin{proof}
 The detailed proof is omitted here for brevity.
\end{proof}

Regarding the convergence properties of VDRW, we would like to make the following two remarks.

\noindent{\bf Remark 1:} As shown in~\cite{pemantle1992vertex}, VRRW enjoys similar convergence properties as VDRW. However, the completely different definition of the visiting function makes the proposed VDRW more suitable for imbalanced networks.

\noindent{\bf Remark 2:} We would like to point out that for real networks, the graph structure hardly meets the convergence conditions in Corollary 3. However, as we will show in the experimental results, short VDRW is still an effective way to extract node-context pairs in learning network representation from imbalanced networks, especially in comparison with state-of-the-art techniques. In Section \ref{case_study}, we further investigate the convergence of VDRW on the real networks.

\section{The Proposed Embedding Framework}
In this section, we present the semi-supervised network representation framework based on Vertex-Diminished Random Walk. Currently most of the existing semi-supervised network representation approaches (e.g.,~\cite{liang2018semi,yang2016revisiting}) aim to learn the network embedding implicitly assuming that the labeled set has roughly equal examples for each class in the network. However, when applied on imbalanced networks, i.e., some classes having significantly more examples (vertices) than the other classes, these approaches will suffer from suboptimal performance. Take Figure \ref{fig:vis}(a) as an example. Karate network has two imbalanced classes where one has 5 labeled examples and the other one has 29 examples. On such an imbalanced network, the embedding learned using~\cite{yang2016revisiting} would make the two classes non-separable from each other, as shown in Figure \ref{fig:imbalance_ex}(a). This is mainly due to inaccurate node-context pairs generated in these methods (i.e., the center node and the context node belong to different classes), which can be greatly improved using the proposed VDRW.

Formally, the imbalanced network embedding problem can be defined as follows.

\begin{problem}{\textbf{(Imbalanced Network Embedding)}}
\begin{description}
\item[Input:] (i) a (directed or undirected) network $G=(V, E)$, and (ii) \textbf{imbalanced class labels} for nodes in $V$
      
\item[Output:]~ a low-dimensional vector representation for each vertex $v \in V$, so that the minority classes are naturally separated from majority class in the embedding feature space.
    \end{description}
\end{problem}

When the label information of only a small portion of the entire network is given, the above problem of imbalanced network embedding would be semi-supervised in nature, as the learned network representation would preserve both the graph structure and the label information in the low-dimensional feature space.

\subsection{Semi-supervised Network Representation Learning}
Suppose that we are given the network $G=(V, E)$ with $ l $ labeled nodes $\{\mathbf{x}_i\}_{i=1}^l$ associated with label $\{\mathbf{y}_i\}_{i=1}^l$\footnote{$\mathbf{y}_i \in \mathbb{R}^C$, where C denotes the number of classes in the network.} and $n-l$ unlabeled nodes $\{\mathbf{x}_i\}_{i=l+1}^{n}$ where $\mathbf{x}_i$ is the feature vector of node $v_i$. The goal is to learn a $d$-dimensional vector $e_i$ for each node $v_i$ using the class label information and the graph structure. The semi-supervised network representation framework can be formulated as an optimization problem with the following loss function:
\begin{equation}
\mathcal{L} = \mathcal{L}_s + \lambda \mathcal{L}_u
\end{equation}
where the first term denotes the prediction loss with respect to the labeled examples, and the second term measures the consistency on the graph. $\lambda$ is a user-defined constant factor that balances between the two terms.

Since the node label can be predicted using both the original features on the node and the embedding features, we use two feed-forward neural network models to train based on the node features and the embedding features, respectively~\cite{yang2016revisiting}, and then concatenate them to predict the label. Therefore, the supervised loss on the labeled nodes can be expressed as follows.
\begin{equation}
\mathcal{L}_s = - \frac{1}{l} \sum_{i=1}^{l}\log{ \frac{\text{exp}([\mathbf{h}^{l_1}(\mathbf{x}_i),\mathbf{h}^{l_2}(\mathbf{e}_i)]^T\mathbf{y}_i)}{\sum_{\mathbf{y'}}\text{exp}([\mathbf{h}^{l_1}(\mathbf{x}_i),\mathbf{h}^{l_2}(\mathbf{e}_i)]^T\mathbf{y'})} }
\end{equation}
where $l_1$ and $l_2$ represent the number of hidden layers in the neural network models using node features and embeddings, respectively. $[\cdot, \cdot]$ concatenates two column vectors. The function $\mathbf{h}(\cdot)$ denotes the output of the neural network model.

Assuming that nearby nodes have the similar embeddings, the graph-based loss function is formulated as follow~\cite{hamilton2017inductive}.
\begin{equation}
\mathcal{L}_u = -\sum_{(v_i, v_c)}(\log{{\sigma(\mathbf{w}_c^T\mathbf{e}_i)}} + k\cdot \mathbb{E}_{v_{n} \sim P_n(v_c)}\log{(\sigma(-\mathbf{w}_{n}^T\mathbf{e}_i))})
\end{equation}
where $(v_i, v_c)$ denotes the node-context pair sampled from the network w.r.t. node $v_i$. $\mathbf{w}_c \in \mathbb{R}^d$ is a vector representation of $v_c$ when $v_c$ acts as the context. Node $v_n$ is randomly selected from a negative sampling distribution $P_n$~\cite{mikolov2013distributed}, and $k$ is the number of negative samples. $\sigma(x) = 1/(1+e^{-x})$ is the sigmoid function.

Of key importance for unsupervised loss is the extraction of node-context pairs in the imbalanced network. As mentioned before, VDRW can capture effectively the node-context pairs for the minority class. Based on that, a \emph{\textbf{Context Sampling}} method is presented to integrate known class-label information with graph structure for extracting node-context pairs. In addition,  mini-batch training method is used in our model training. During the embedding learning, a batch of nodes is randomly selected as the initial points $\mathcal{S}$ to extract node-context pairs at each iteration. Majority class has the higher probability to be selected according to the imbalanced distribution than minority class. Thus, \emph{\textbf{Balanced-batch Sampling}} is introduced to balance those pairs with respect to class-label in model training.

The training process in our network representation learning framework is illustrated in Algorithm \ref{MT}. It is given the graph $G$ with node features $\mathbf{x}_{1:n}$ and partial labels $y_{1:l}$, and the training model parameters (batch iterations $T1$ and $T2$, the negative sampling size $k$, embedding size $d$) as input, and outputs the embedding vector $\mathbf{e}_i$ for each node $v_i$. Lines 1-8 demonstrate the training process in learning the embedding vector for each node in the imbalanced network. Lines 9-12 correspond to the process of predicting the label using node feature $\mathbf{x}_i$ and embedding vector $\mathbf{e}_i$. And stochastic gradient descent (SGD) is adopted to train our model.

\begin{algorithm}[htp]
\caption{\emph{\textbf{Model Training}}}\label{MT}
\leftline{\textbf{Input:} Graph $G$ with features $\mathbf{x}_{1:n}$ and labels $y_{1:l}$}
\hspace*{1cm} \leftline{batch iterations $T1$ and $T2$}
\hspace*{1cm} \leftline{negative sampling size $k$}
\hspace*{1cm} \leftline{embedding size $d$}
\leftline{\textbf{Output:} Embedding vector $\{\mathbf{e}_i\}_{i=1}^{n}$}
\begin{algorithmic}[1]
\For{$t$ $\gets$ 1 to $T1$}
\State Initialize $\mathcal{S}$ using \emph{\textbf{Balanced-batch Sampling}};
\For{$s$ $\gets$ 1 to $|\mathcal{S}|$}
\State Sample positive pairs using \emph{\textbf{Context Sampling}};
\State Sample $k$ negative pairs;
\EndFor
\State{\textbf{end for}}
\State Update the model parameters in Eq. (8) using SGD;
\EndFor
\State{\textbf{end for}}
\For{$t$ $\gets$ 1 to $T2$}
\State Sample a batch of $\{\mathbf{x}_i, y_i\}$ from labeled instances;
\State Update the model parameters in Eq. (7) using SGD;
\EndFor
\State{\textbf{end for}}
\State Return the embedding vector $\mathbf{e}_i \in \mathbb{R}^d$ for each node.
\end{algorithmic}
\end{algorithm}

\subsection{Context Sampling}
In node-context pairs extraction, DeepWalk~\cite{perozzi2014deepwalk} has proven the effectiveness of random walk in exploring the context for each node in the network. The core idea in random walk is to choose one neighbor based on the transition probabilities. As discussed in Section \ref{Vertex-Diminished Random Walk}, random walk might not capture the high-quality node-context pairs, especially for the nodes within minority class. Thus, VDRW is introduced to explore the context for nodes by adjusting the transition probabilities with respect to their visited times. Based on the proposed VDRW, we present a \emph{\textbf{Context Sampling}} method integrating label information with network topological structure.

In semi-supervised learning, it is implicitly assumed that examples from the same class would be close to each other in the feature space. That is the reason why labeled nodes can naturally be the contexts for each other if they have identical labels. Planetoid~\cite{yang2016revisiting} also exploited this assumption. However, they extracted node-context pairs using labels and graph structure separably. That is, context is either determined by random walk or by labels, but not both. Here we adopt a jumping strategy in which one node may jump directly to another node during random walk if they have the same labels. That would encourage one node to explore the context though they are not close in the network. To some extent, it enlarges the size of the neighborhood by using known label information.

The \emph{\textbf{Context Sampling}} method for extracting node-context pairs is illustrated in Algorithm \ref{context_sampling}. If one node is labeled, it can visit another node with the same label directly with probability $r$ ($0<r<1$). With probability $1-r$, it selects the neighbors based on the transition matrix. After each step, the transition matrix is updated using VDRW. The node-context pairs can be extracted from the walking path as did in DeepWalk~\cite{perozzi2014deepwalk}.
\begin{algorithm}[htp]
\caption{\emph{\textbf{Context Sampling}}}\label{context_sampling}
\leftline{\textbf{Input:} Graph $G$ with transition matrix $\mathbf{R}$, labels $y_{1:l}$}
\hspace*{1cm} \leftline{initial node $v_s$}
\hspace*{1cm} \leftline{jumping probability $r$}
\hspace*{1cm} \leftline{length of walking sequences $T$}
\leftline{\textbf{Output:} Node-context pairs $(v_i, v_c)$}
\begin{algorithmic}[1]
\State \textbf{Initialize:} Add $v_s$ to walking path $p$;
\For{$t=$ 0 to $T-1$}
\If{Node is labeled}
\State With $r$, select one node $v$ with the same label;
\State With $1-r$, select its neighbor $v$ based on $\mathbf{R}$;
\Else
\State Select its neighbor $v$ based on $\mathbf{R}$;
\EndIf
\State{\textbf{end if}}
\State Update $\mathbf{R}$ using VDRW;
\State Update the walking path $p$;
\EndFor
\State{\textbf{end for}}
\State Return the node-context pairs sampled from path $p$.
\end{algorithmic}
\end{algorithm}

\subsection{Balanced-batch Sampling}
Mini-batch training is commonly used in building neural network models~\cite{yang2016revisiting,liang2018semi} assuming that the training instances are class-balanced. In our framework, however, the number of node-context pairs in different classes would be typically imbalanced if we randomly select a batch of nodes as the initial nodes. In other words, the imbalance characteristic is preserved in those mini-batches even though we could find node-context pairs correctly. To avoid this issue, we adopt a simple \emph{\textbf{Balanced-batch Sampling}} method for mini-batch selection at each iteration.

Considering the binary imbalanced learning as an example, there are three different sets of nodes on the training model: labeled majority $S_{maj}$, labeled minority $S_{min}$ and unlabeled instances $S_{u}$. In most cases, the batch size is significantly larger than the number of labeled instances as only a very small portion of nodes are labeled.
Balanced-batch sampling randomly selects a subset $S'_{maj}$ from the labeled majority class such that $|S'_{maj}|=|S_{min}|$, and then sampled a subset $S'_{u}$ from the unlabeled data. Consequently, the set given by \emph{\textbf{Balanced-batch Sampling}} $S = S_{min} \cup S'_{maj} \cup S'_{u}$ will be used as initial nodes in model training. 

It is close to the under-sampling method which reduces the instance to balance data distribution. In general, random under-sampling may cause the training model to miss important details related to the majority class~\cite{he2009learning}. Nevertheless, it would be alleviated in our mini-batch training because it considers the entire graph nodes via iterative training. Note that in mini-batch model training, the examples in the minority class might be adopted at each iteration by combining with randomly selected examples in the majority class if they are less than the batch size $T1$.

\section{Experiments}
In this section, we conduct the experiments on several real-world networks to validate the effectiveness of the proposed \verde{} by comparing to several state-of-the-art methods. Vertex-Diminished Random Walk (VDRW) is proposed to extract the node-context pairs for our imbalanced network embedding framework, which is denoted as {\it ImVerde-VDRW}. Additionally, Vertex-Reinforced Random Walk (VRRW)~\cite{pemantle1992vertex} is also considered in our framework (denoted as {\it ImVerde-VRRW}) for comparison.

In particular, we focus on the following questions: (1) How effective is short VDRW for exploring the context on the imbalanced networks compared with short random walk and VRRW? (2) Compared to other state-of-the-art methods, how effective is the \verde{} framework for learning network representation from imbalanced data? (3) What are the impacts of class-imbalanced distribution for the \verde{}?

\subsection{Setup}
\textbf{Data sets:}
We use several node classification data sets to validate the effectiveness of the proposed method, including Karate, Cora, Citeseer, Pubmed and NTSB data sets. Table \ref{tab:data} listed the detailed data description. The model effectiveness on those data sets can be evaluated by predicting the correct label for nodes in the minority class.
\begin{itemize}
\item Karate network~\cite{zachary1977information}. It is a social network of the karate club, which represents the social relationship of 34 members in the karate club. Karate network has 34 nodes and 156 edges. To validate the effectiveness of VDRW on imbalanced data, We consider it as a binary imbalanced network as shown in Figure \ref{fig:vis}(a) where the minority class contains only 5 nodes and the other 29 nodes belong to the majority class.

\item Cora~\cite{mccallum2000automating}, Citeseer~\cite{giles1998citeseer} and Pubmed~\cite{namata2012query} networks. All of them are the citation networks. For those data sets, each node corresponds to one scientific publication and the edge represents the citation relationship between two publications. And each publication is described by a sparse bag-of-words feature vector. Originally, there are several classes for every data set where each class has the roughly equal number of examples (nodes). In our experiments, we reconstruct them for binary imbalanced network analysis. Specifically, we select 20 nodes from one class and 120 nodes from the rest classes as the labeled training examples. Besides, 1000 nodes are chosen as the test examples and others are unlabeled training examples. Take Cora data set as an example, seven imbalanced binary data sets are possible to be reconstructed considering choosing one of them as the minority class (marked as Cora\_1, $\cdots$, Cora\_7, respectively). The imbalance ratio of the number of examples in minority class over that in majority class is 20:120 for each reconstructed Cora data set.

\item NTSB network. It is built from the historical accident reports published in National Transportation Safety Board database\footnote{\url{https://www.ntsb.gov/Pages/default.aspx}}. Those reports can be naturally modeled as a network in which each node represents one accident report and each edge reflects the similarity between two reports. It contains 1473 nodes and 11784 edges where each node is associated with a 4424-dimensional TF-IDF feature vector~\cite{salton1988term} extracted from the accident report. For NTSB data set, the imbalance ratio is 5:5:5:200. Therefore, it is a significantly imbalanced network with 4 classes. The number of labeled training instances in the majority class is 40 times more than that of each minority class. And there are 500 test instances. This data set is used to validate the effectiveness of the proposed method on multi-class classification from imbalanced data.
\end{itemize}

\begin{table}[htp]
\centering
  \caption{Data statistics}
  \label{tab:data}
  \begin{tabular}{lrrrrr}
    \hline
    Dataset & Karate & Cora & Citeseer & Pubmed & NTSB \\
    \hline
    Nodes & 34 & 2,708 & 3,327 & 19,717 & 1,473     \\
    Edges & 156 & 5,429 & 4,732 & 44,338 & 11,784 \\
    Classes & 2 & 7 & 6 & 3 & 4\\
    Features & --- & 1,433 & 3,703 & 500 & 4,424 \\
  \hline
\end{tabular}
\end{table}

\textbf{Baselines:} Four baseline methods are adopted in our experiments to validate the effectiveness of the proposed approach.
\begin{itemize}
\item DeepWalk~\cite{perozzi2014deepwalk}. It is an unsupervised network embedding method which first considered to use truncated random walk to explore the context as word embedding models. The learned node representations encoded the graph topological structure in a continuous vector space.

\item node2vec~\cite{grover2016node2vec}. It is a semi-supervised model for learning node representation in networks. In order to smoothly interpolate between BFS and DFS, it designed a biased random walk for neighborhood definition where two parameters $p$ and $q$ were adopted to balance the depth and width of node neighborhood. And these parameters would be learned using the labeled training data in a semi-supervised fashion.

\item Planetoid~\cite{yang2016revisiting}. It is also a semi-supervised learning frame- work where each node's class label and neighborhood context would be jointly predicted. It used both random walk on the graph and node label information to define the neighborhood context based on the assumption that nearby nodes or nodes with the same class label have the similar embeddings in the feature space. The authors provided both transductive and inductive versions of Planetoid (Planetoid-T and Planetoid-I) for learning node representation. In this paper, we only consider the transductive version (denoted as Planetoid in our experiments instead of Planetoid-T in \cite{yang2016revisiting}). The inductive \verde{} can be modeled using the similar strategy as Planetoid-I.

\item  SMOTE\cite{chawla2002smote}. It is an over-sampling approach for imbalanced data, in which the synthetic samples are created between each minority instance and its nearest neighbors. Therefore, we use this method to simply balance the data distribution on the original feature space. It assumed that the data are independent and identically distributed without considering any graph properties.

\end{itemize}

\subsection{Case Study: Zachary's Karate Network}
\label{case_study}
\begin{figure}
\centering
\subfigure[Short walking path accuracy]{
\begin{minipage}{0.45\linewidth}
\centering
\includegraphics[width=1.7in]{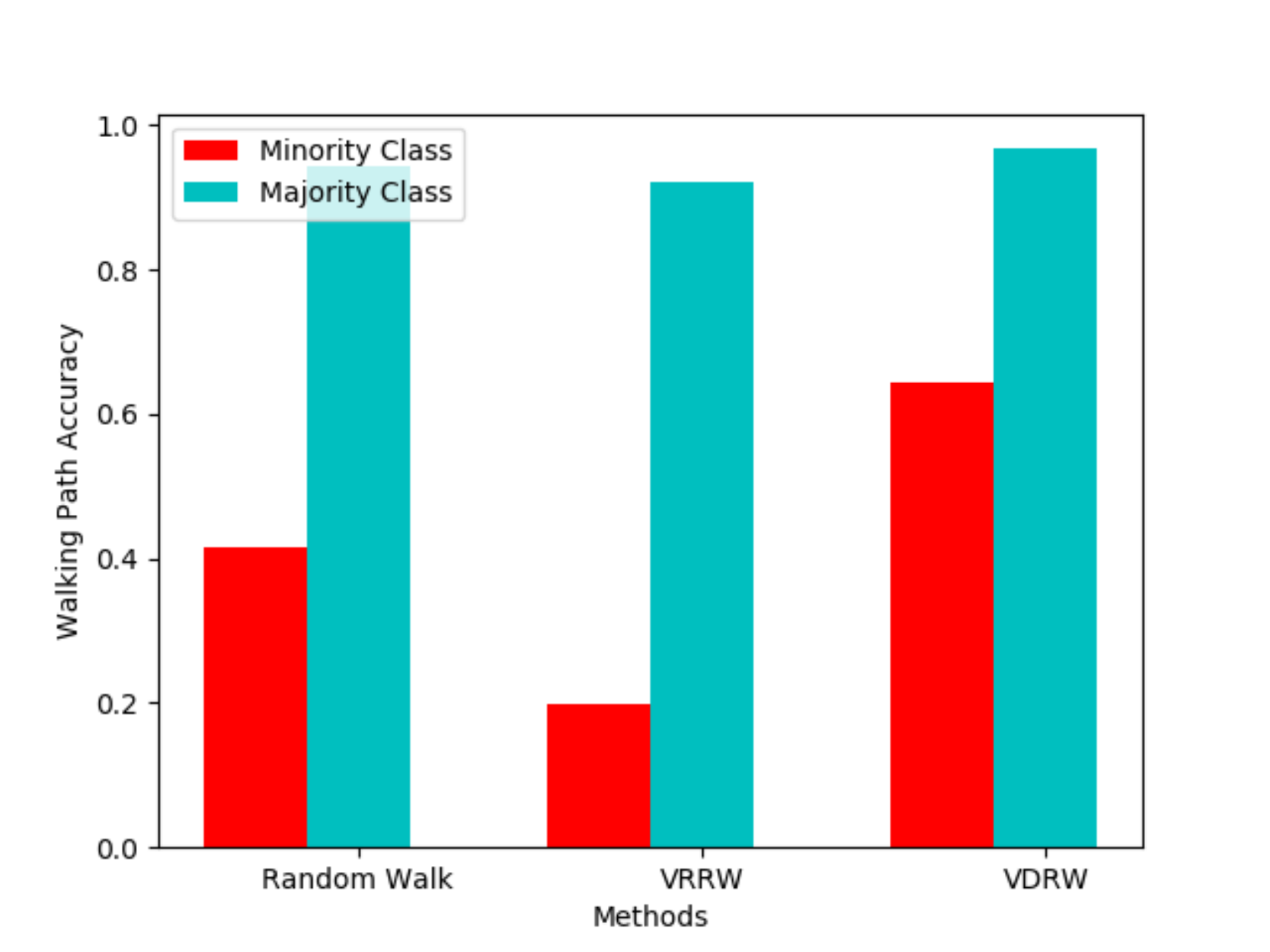}
\end{minipage}
}
\subfigure[Convergence analysis]{
\begin{minipage}{0.45\linewidth}
\centering
\includegraphics[width=1.7in]{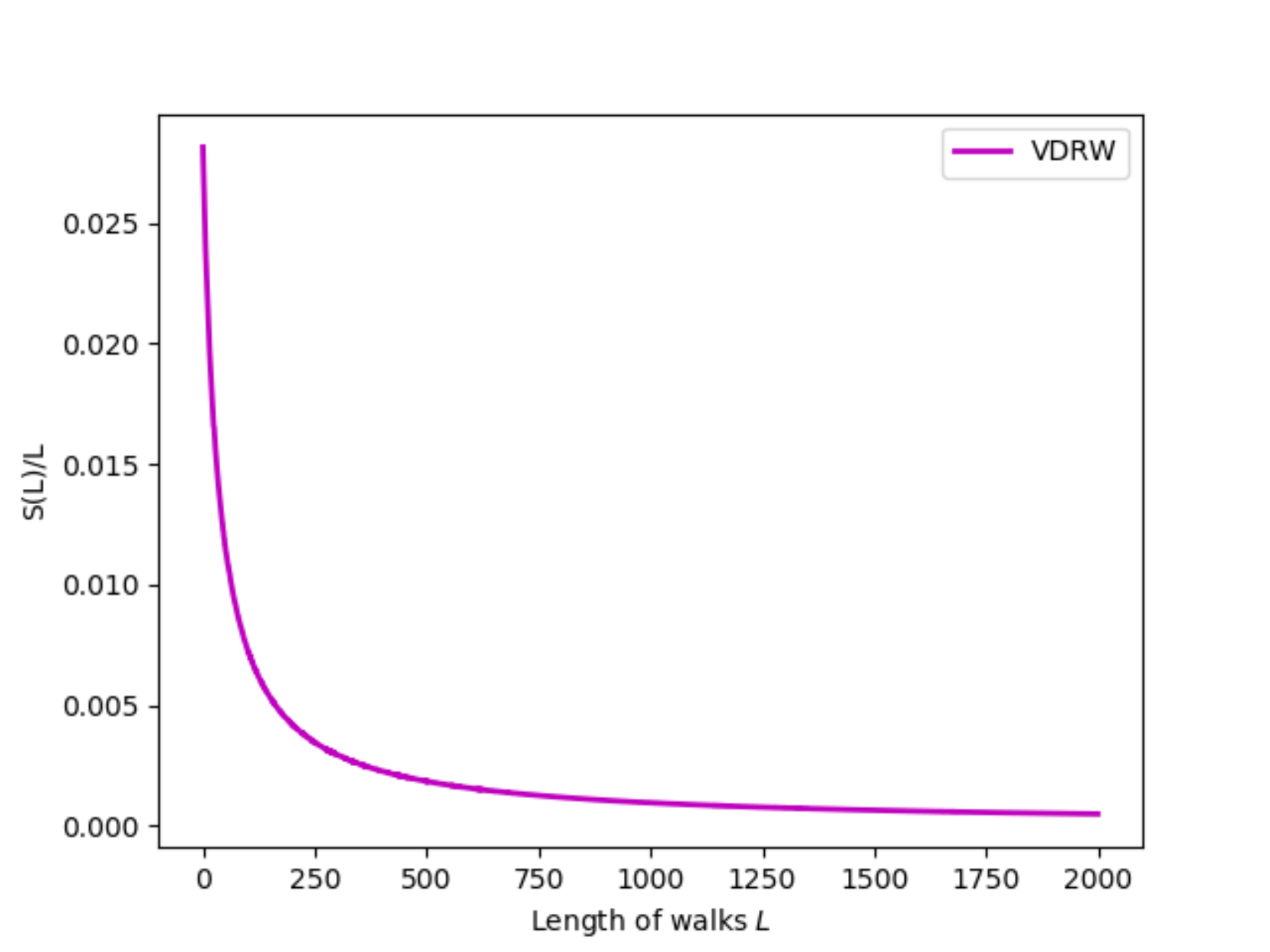}
\end{minipage}
}
\caption{The analysis of VDRW on Karate network}\label{vdrw_poperty}
\vspace{-6mm}
\end{figure}

 In Section \ref{Vertex-Diminished Random Walk}, we observed that VDRW encourages the nodes to be visited with similar frequency. This characterization would force short VDRW to walk within the same class. We conduct a case study on Zachary's Karate network~\cite{zachary1977information} to validate the conclusion empirically. The strategies including random walk, VRRW and VDRW are used to walk for short steps starting from each node. The walking length and walking times are set as 10 and 100, respectively. We statistic the walking path accuracy, which defines how many nodes in the walking path are located within the same class. Figure \ref{vdrw_poperty}(a) gives the average walking accuracy in both minority and majority class for those methods. It can be seen that VDRW improves the walking accuracy in both classes. That is, it encourages then nodes to walk within the same class. As the node-context pairs are extracted from the walking path in \verde{}, the high-quality node-context pairs would improve the embedding performance for imbalanced networks. The convergence of VDRW is also evaluated on Karate network. We statistic the difference of the visited times $\mathbf{S}(L)/L$ after each step. As shown in Figure \ref{vdrw_poperty}(b), the difference keeps decreasing with respect to the walking steps $L$. Additionally, the node representations learned by \verde{} and Planetoid are visualized in the 2-dimensional space using t-SNE as shown in Figure \ref{fig:imbalance_ex}.

\subsection{Node Classification}
We conduct the node classification experiments on four real networks to evaluate the proposed framework \verde{}. In VDRW, we select the parameters $\alpha = 0.7$ and $r=0.2$, which is discussed in details later. And we use the same random walk parameters for all the used methods: walking length $T=10$, negative samples $k=10$, embedding size $d=50$. The logistic regression classifier is used to evaluate the node representations learned from DeepWalk, node2vec and SMOTE in our experiments.
\begin{figure*}[htp]
\centering
\subfigure[Cora]{
\begin{minipage}{0.3\linewidth}
\centering
\includegraphics[width=2.4in]{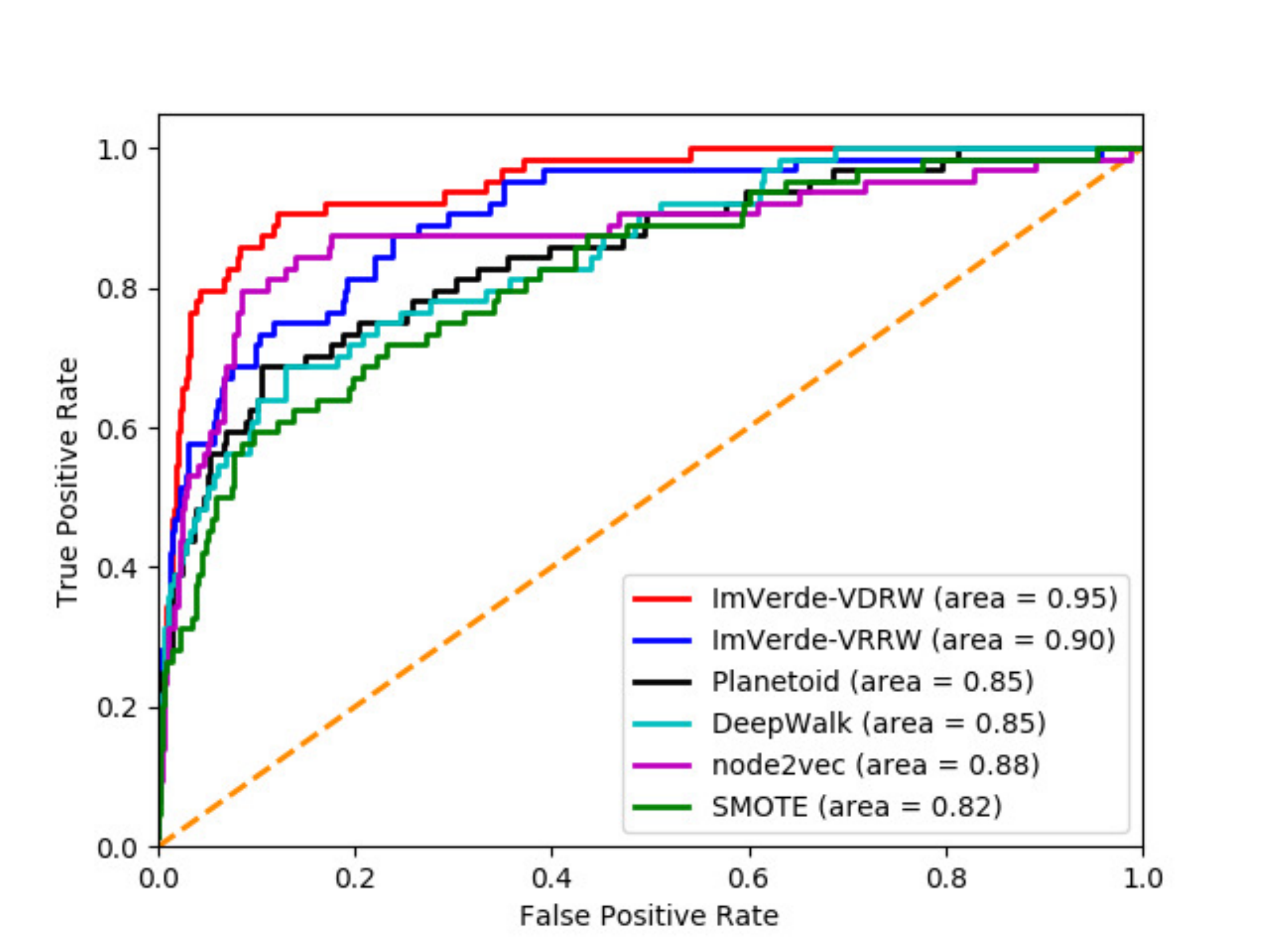}
\end{minipage}
}
\subfigure[Citeseer]{
\begin{minipage}{0.3\linewidth}
\centering
\includegraphics[width=2.4in]{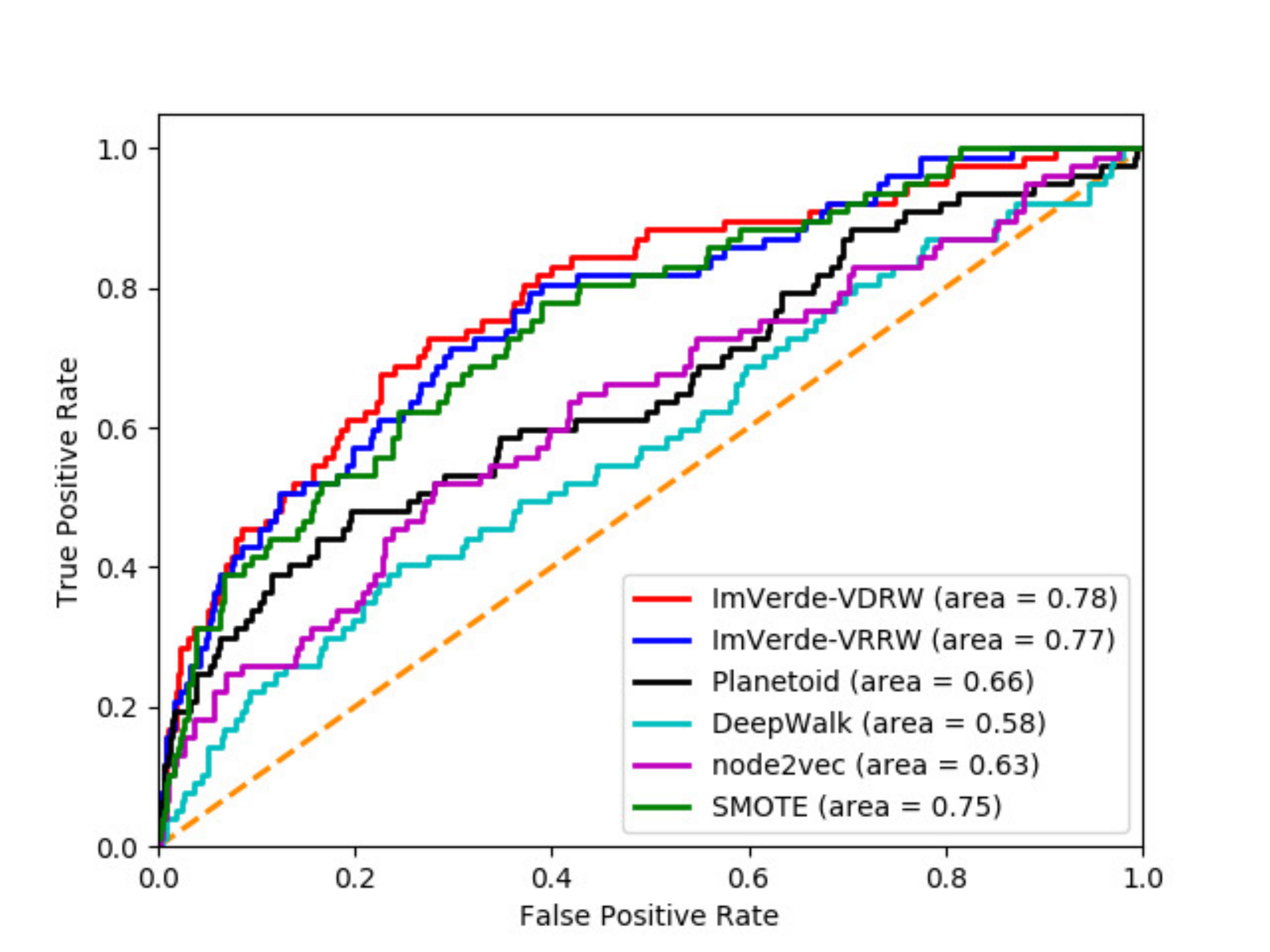}
\end{minipage}
}
\subfigure[Pubmed]{
\begin{minipage}{0.3\linewidth}
\centering
\includegraphics[width=2.4in]{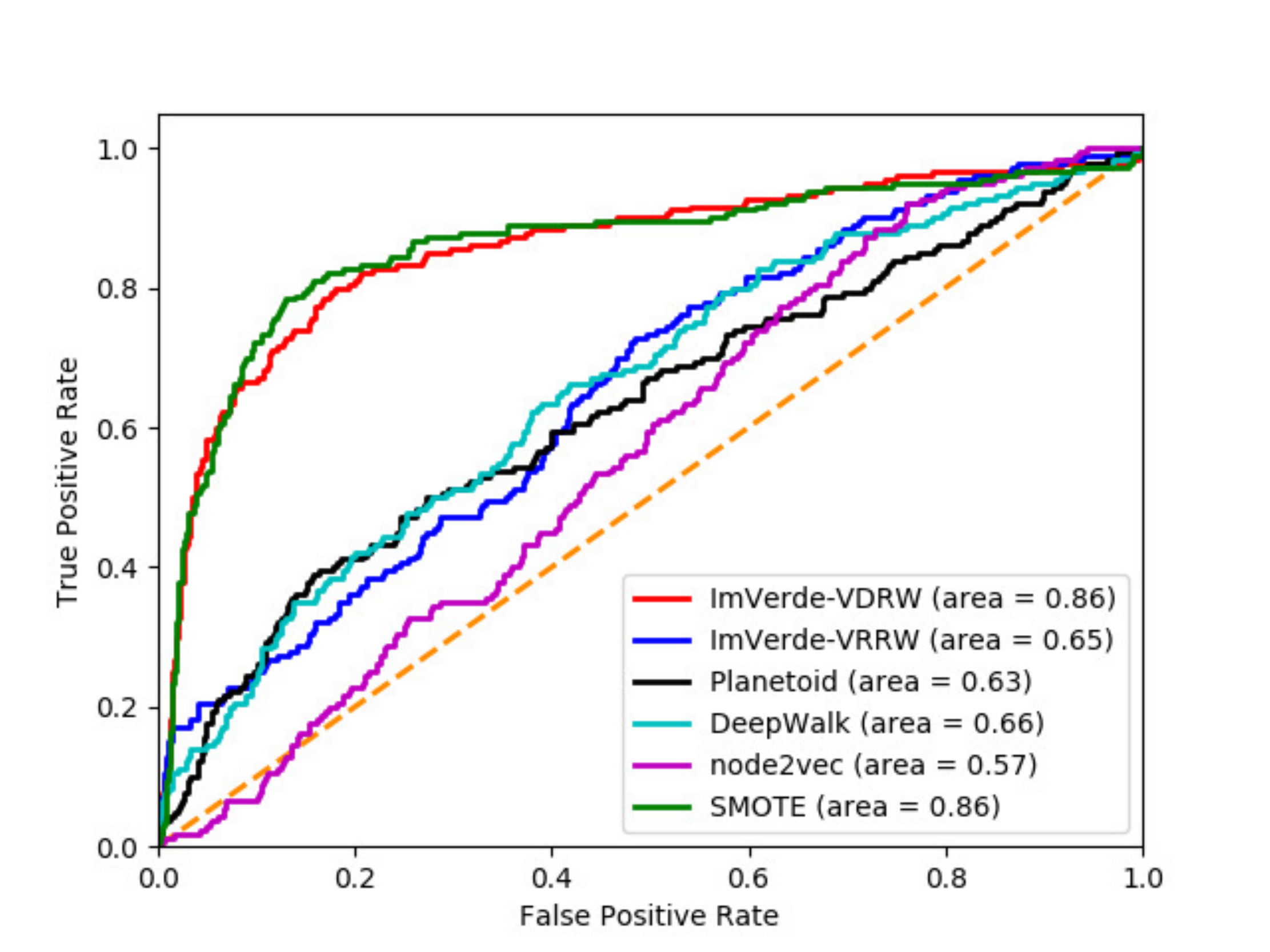}
\end{minipage}
}
\caption{The ROC curve and AUC value for node classification on Cora, Citeseer and Pubmed data sets (best seen in color)}\label{node_classification}
\vspace{-6mm}
\end{figure*}

\begin{table}[htp]
\caption{Average precision for the binary node classification (The best results are indicated in bold)}
\label{AP_cora}
\centering
\scriptsize
\begin{tabular}{cccccc}
\hline
 & Planetoid & DeepWalk & node2vec & SMOTE & {\it ImVerde-VDRW}\\
\hline
Cora\_1 & 0.639 & 0.650 & 0.612 & 0.473  & \textbf{0.720} \\
Cora\_2 & 0.749 & 0.778 & 0.761 & 0.664  & \textbf{0.852} \\
Cora\_3 & 0.900 & 0.863 & 0.840 & 0.707  & \textbf{0.910} \\
Cora\_4 & 0.783 & 0.657 & 0.761 & 0.723  & \textbf{0.826} \\
Cora\_5 & 0.637 & 0.610 & 0.471 & 0.641 & \textbf{0.769} \\
Cora\_6 & 0.557 & 0.722 & 0.505 & 0.584 & \textbf{0.805}\\
Cora\_7 & 0.501 & 0.474 & 0.503 & 0.404  & \textbf{0.657} \\
\hline
Citeseer\_1 & 0.238 & 0.116 & 0.169 & 0.252 & \textbf{0.304} \\
Citeseer\_2 & 0.396 & 0.225 & 0.250 & \textbf{0.550} & 0.513 \\
Citeseer\_3 & 0.651 & 0.616 & \textbf{0.693} & 0.605 & 0.666 \\
Citeseer\_4 & 0.608 & 0.352 & 0.415 & 0.729 & \textbf{0.732} \\
Citeseer\_5 & 0.706 & 0.512 & 0.610 & \textbf{0.790} & 0.785 \\
Citeseer\_6 & 0.591 & 0.310 & 0.432 & 0.644 & \textbf{0.651} \\
\hline
Pubmed\_1 & 0.298 & 0.335 & 0.203 & 0.631 & \textbf{0.631} \\
Pubmed\_2 & 0.454 & 0.632 & 0.497 & \textbf{0.669} & 0.663 \\
Pubmed\_3 & 0.489 & \textbf{0.736} & 0.462 & 0.685 & 0.653 \\
\hline
\end{tabular}
\end{table}
\vspace{-3mm}

\begin{table}[htp]
\centering
  \caption{Multi-class text classification on NTSB data set}
  \label{tab:ntsb}
  \begin{tabular}{cc}
    \hline
     & Accuracy \\
    \hline
    Planetoid & 0.311\\
    DeepWalk     & 0.458 \\
    node2vec &  0.408\\
    {\it ImVerde-VRRW}  & 0.416\\
    {\it ImVerde-VDRW} & \textbf{0.460}\\
  \hline
\end{tabular}
\end{table}

Figure \ref{node_classification}(a) gives the Receiver Operating Characteristic (ROC) curve and AUC value for predicting the minority class on Cora\_7, which means that the 7-th class is selected as the minority while other classes are the majority on Cora. We observe the similar results on other reconstructed binary data sets of Cora, but omit them due to the limited space. Figure \ref{node_classification}(b) and Figure \ref{node_classification}(c) give the ROC curves and AUC values on Citeseer\_6 and Pubmed\_1, respectively. More specifically, Table \ref{AP_cora} lists the average precisions on Cora, Citeseer and Pubmed data sets where the best results are highlighted in bold. Besides, we conduct the multi-class node classification on NTSB data set with class-imbalanced labeled training instances. The multi-classification accuracy on NTSB data set is given in Table \ref{tab:ntsb}. It shows that {\it ImVerde-VDRW} outperforms {\it ImVerde-VRRW} in these data sets. Short VDRW encourages the node-context pairs to be extracted within the same class. On the contrary, short VRRW would force one node to visit other high-degree nodes. The quality of node-context pairs is crucial to the random walk based embedding approaches when the labeled data is class-imbalanced. 

We achieve the significant improvement over DeepWalk on Cora, Citeseer and Pubmed. DeepWalk\footnote{\url{https://github.com/phanein/deepwalk}} adopted short random walk to extract the context for each node from the network.
Node2vec\footnote{\url{https://snap.stanford.edu/node2vec}} designed another random walk strategy to interpolate between BFS and DFS. It stated that the likelihood of immediately revisiting a node decreases. That idea is similar to VDRW which decreases the transition probability each step to explore the neighborhood. The short random walk in DeepWalk can be seen as a special case for both node2vec and VDRW ($p=q=1$ for node2vec, $\alpha = 1$ for VDRW).
But there are some significant differences between them. First, the affected area is different. When node $v_j$ is visited from $v_i$, node2vec considers updating the probability of links $v_j$ will visit next time. In contrast, the probability of any transition to $v_j$ would decrease in VDRW. Second, the transition probability fluctuates temporarily in node2vec, whereas VDRW in \verde{} preserves the diminished probability for the whole node-context extraction process.
Planetoid\footnote{\url{https://github.com/kimiyoung/planetoid}} is a semi-supervised learning framework which used both random walk and labels to extract node-context. But it does not perform well on imbalanced data classification due to the node-context pairs extracted using random walk. On the other hand, SMOTE\footnote{\url{https://github.com/scikit-learn-contrib/imbalanced-learn}} simply balanced the data distribution using attributes without graph information. Thus the node attributes have a great impact on its classification performance. It achieves the worst performance on Cora data set, but gets excellent classification results on other data sets. In such cases, the embedding representation learned from {\it ImVerde-VDRW} is still competitive.

\subsection{Imbalance Analysis}
\begin{figure}
\centering
\subfigure[The imbalance ratio analysis]{
\begin{minipage}{0.45\linewidth}
\centering
\includegraphics[width=1.7in]{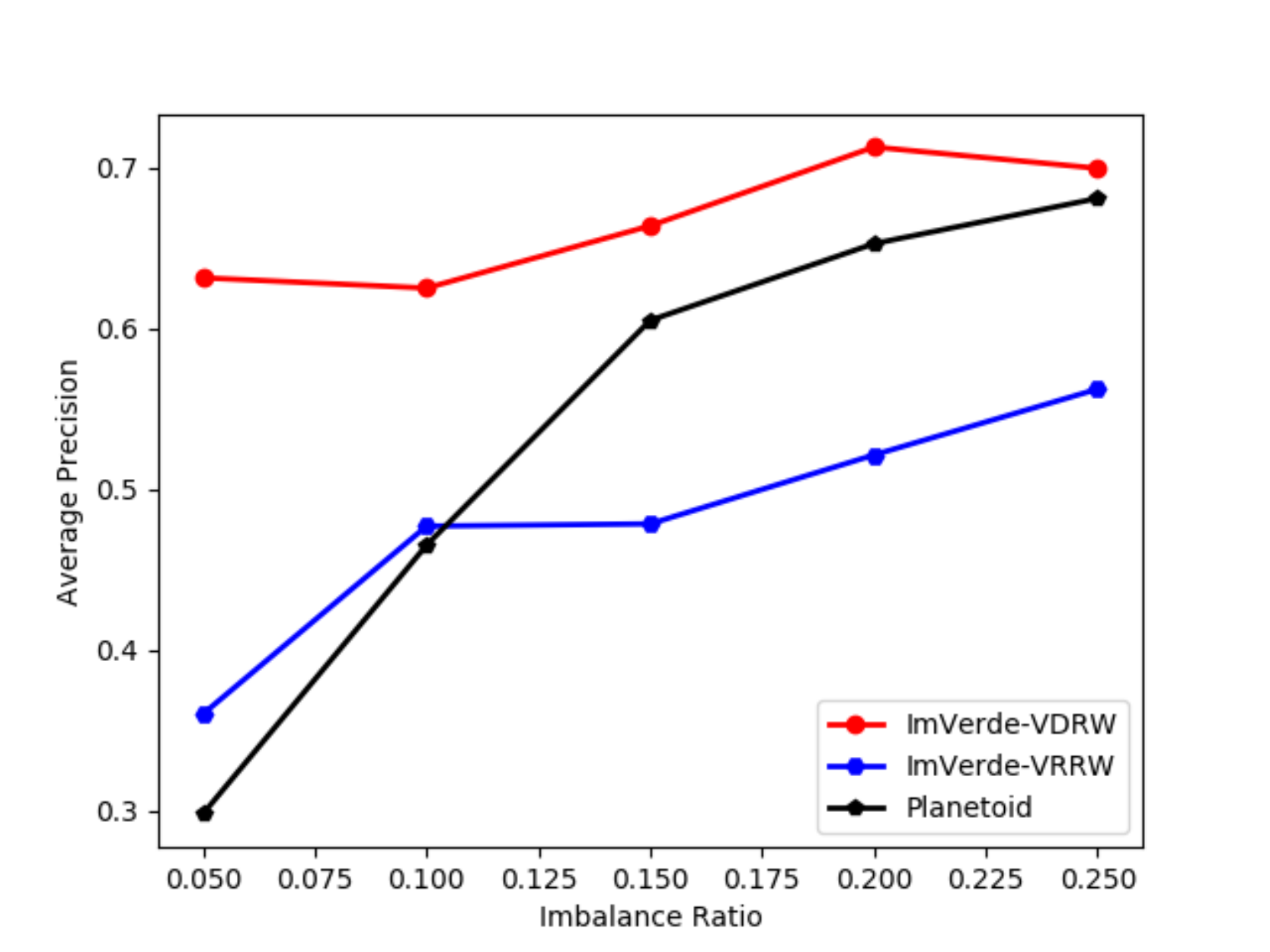}
\end{minipage}
}
\subfigure[Parameter sensitivity analysis]{
\begin{minipage}{0.45\linewidth}
\centering
\includegraphics[width=1.7in]{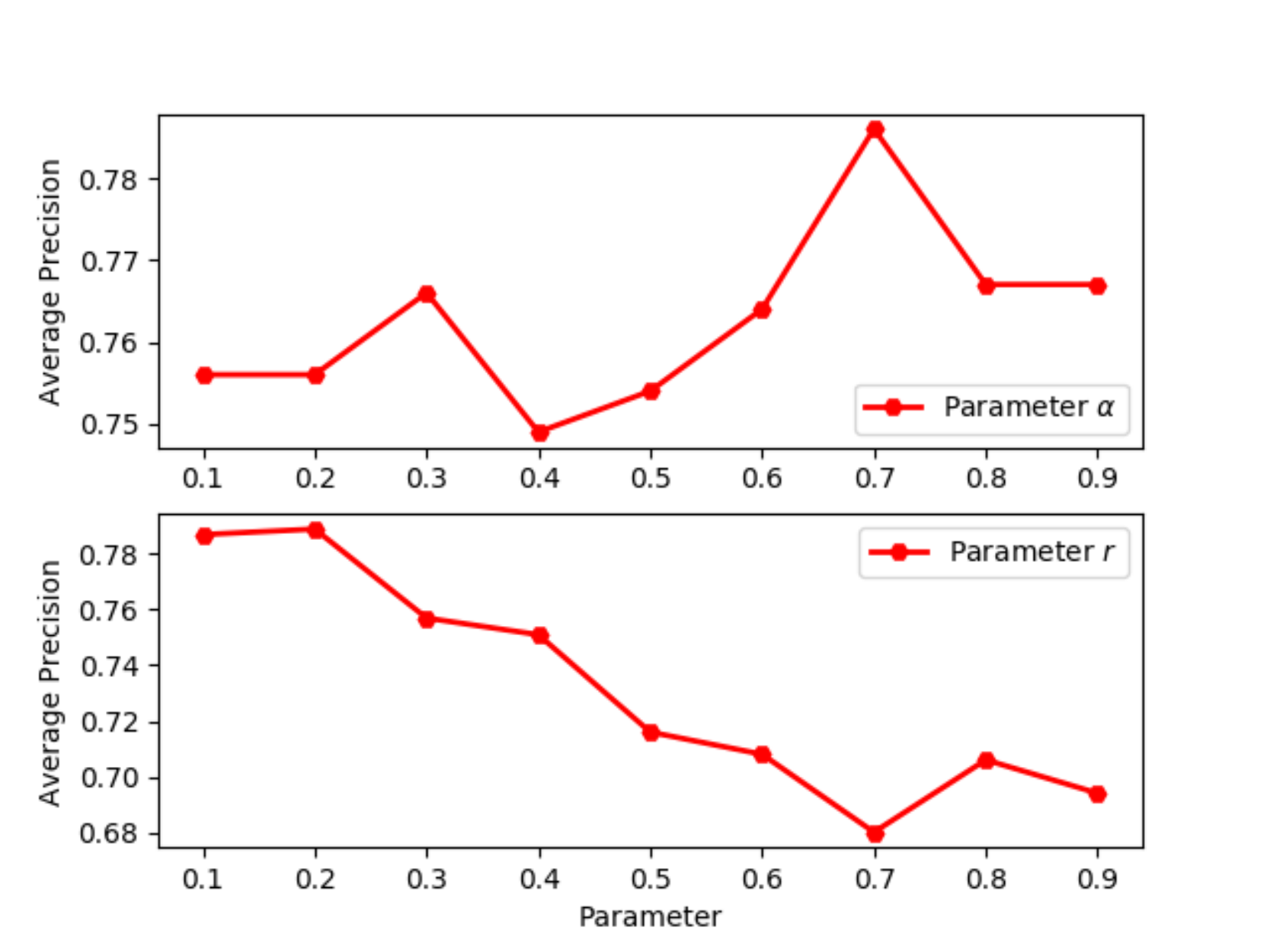}
\end{minipage}
}
\caption{Model analysis}\label{model_analysis}
\vspace{-6mm}
\end{figure}

Since we aim to learn the network representation from imbalanced data, the imbalance ratio impacts on the embedding performance for our framework. We consider three semi-supervised learning frameworks: Planetoid, {\it ImVerde-VRRW} and {\it ImVerde-VDRW}.
Figure \ref{model_analysis}(a) gives the binary node classification result on Pubmed\_1 with different imbalance ratio using different frameworks. When the imbalance ratio ranges from 0.050 to 0.250, it shows that the average precision of {\it ImVerde-VDRW} is slightly affected compared to Planetoid and {\it ImVerde-VRRW}. And {\it ImVerde-VDRW} outperforms other two methods when the network is extremely class-imbalanced. The classification performance of Planetoid is heavily relying on the imbalance ratio of labeled training data because it implicitly assumes that the labeled instances have the approximately balanced distribution.

\subsection{Parameter Sensitivity}
There are two crucial hyper-parameters ($\alpha$ and $r$) in our \emph{\textbf{Context Sampling}} strategy of \verde{} where $\alpha$ ($0 < \alpha < 1$) controls how the transition probability decreases w.r.t. visited times in VDRW and $r$ is the probability of jumping to another node with the same label in \emph{\textbf{Context Sampling}}. Take Cora\_1 data set as an example, we investigate the node-classification performance with different parameter values. Figure \ref{model_analysis}(b) shows the average precision of the proposed framework using different parameter values, where we fix one parameter when investigating the classification performance w.r.t the other one. It can be observed that both $\alpha$ and $r$ affect the classification performance of \verde{}.

\section{Conclusion}
In this paper, we present a novel Vertex-Diminished Random Walk model, which reduces the transition probability to one node each time it is visited. This property makes it particularly suitable for analyzing imbalanced networks, as it encourages the random particle to walk within the same class, thus generating more accurate node-context pairs for the following network embedding task. Then based on VDRW, we propose \verde{}, a semi-supervised network representation learning framework for imbalanced networks. In this framework, we use VDRW and the label information to capture node-context pairs by assuming nodes with the same context or label have the similar embedding feature. Furthermore, we adopt a simple under-sampling method to balance those pairs from different classes. We compare \verde{} with state-of-the-art techniques on several imbalanced networks. The experimental results demonstrate the effectiveness of the proposed framework on node classification, especially when the labeled data is extremely class-imbalanced.

\section*{Acknowledgment}
This work is supported by the United States Air Force and DARPA
under contract number FA8750-17-C-0153, National Science Foundation
under Grant No. IIS-1552654, Grant No. IIS-1813464 and
Grant No. CNS-1629888, NASA under Grant No. NNX17AJ86A, the U.S. Department of Homeland Security under Grant Award Number 17STQAC00001-02-00, and an IBM
Faculty Award. The views and conclusions are those of the authors
and should not be interpreted as representing the official policies
of the funding agencies or the government.

\end{document}